\documentclass[conference]{IEEEtran}
\ifCLASSINFOpdf
\else
\fi

\usepackage{amsmath,amsthm}
\theoremstyle{problem}

\usepackage{amsfonts}
\usepackage{amsmath,amsthm}
\usepackage{graphicx,epstopdf,epsfig,multirow,epic,bm}
\usepackage{color}
\usepackage{multicol}
\usepackage{CJK}
\usepackage{makeidx,showidx}
\usepackage{ifthen}
\usepackage{mathrsfs}
\usepackage{colortbl}
\usepackage{multicol}
\usepackage{latexsym}
\usepackage{amssymb}
\usepackage{algorithm}
\usepackage{algorithmic}
\usepackage{longtable}
\usepackage{array}

\usepackage{paralist}

\theoremstyle{plain}
\newtheorem{thm}{Theorem}

\theoremstyle{definition}
\newtheorem{defn}{Definition}
\theoremstyle{definition}
\newtheorem{rem}{Remark}

\theoremstyle{definition}

\theoremstyle{plain}
\newtheorem{conj}[thm]{Conjecture}

\begin{document}

\begin{CJK}{GBK}{song}
%
\title{Graph Theory Towards New Graphical Passwords In Information Networks}


\author{\IEEEauthorblockN{Bing \textsc{Yao}$^{\ddagger,1,3}$\quad Hui \textsc{Sun}$^{1}$\quad  Hongyu \textsc{Wang}$^{2}$\quad  Jing \textsc{Su}$^{1,2}$\quad  Jin \textsc{Xu}$^{2}$}
\IEEEauthorblockA{{1} College of Mathematics and Statistics,
 Northwest Normal University, Lanzhou,  730070,  CHINA}
\IEEEauthorblockA{{2} School of Electronics Engineering and Computer Science,
Peking University, Beijing, 100871,  CHINA}
\IEEEauthorblockA{ {3} School of Electronics and Information Engineering, Lanzhou Jiaotong University, Lanzhou, 730070, CHINA\\
$^\ddagger$ Corresponding author's email: yybb918@163.com}

\thanks{Manuscript received xxxx 1, 2017; revised xxxx, 2017.
Corresponding author's email: yybb918@163.com.}}


%


\maketitle

\begin{abstract}
Graphical passwords (GPWs) have been studied over 20 years. We are motivated from QR codes that can be considered GPWs, are successfully applied in today's world. However, one need such GPWs that can be use conveniently and have higher level security in information networks. We aim to GPWs for mobile devices with touch screen. No more researching papers of GPWs by means of techniques of graph theory that are already used in a wide range of scientific areas, especially, dynamic networks. Our investigation is devoted to designing new type of graphical passwords by methods of graph theory, such as various topological structures and colorings/labellings. We develop the idea of ``topological structure plus number theory'' in maximal planar graphs, and provide new techniques for designing new type of GPWs, and furthermore we do theoretical analysis on these new type of GPWs.\\[4pt]
\end{abstract}
\textbf{\emph{Keywords}---Graphical password; coloring; labelling; Kempe equivalence; trees; operation.}


%
\IEEEpeerreviewmaketitle

\section{Introduction}

With the rapid development of intelligence, the password becomes the guarantee of normal operation of intelligent mechanism such as networks. The multivariate, multi-formalized and widely applied cryptographs have become an urgent need, and the production of various passwords also face to hard attackers. Classical studies of the password problem going back over 35 years (Morris and Thompson, 1979; Feldmeier and Karn, 1990; Klein, 1990) have shown that, as a result, human users tend to choose and handle alphanumeric passwords very insecurely \cite{Wiedenbeck-Waters-Birget-Brodskiy-Memon-2005}. The following password problem is well known in the security community:

\textbf{The password problem \cite{Wiedenbeck-Waters-Birget-Brodskiy-Memon-2005}:} (1) Passwords should be easy to remember, and the user authentication protocol should be executable quickly and easily by humans.

(2) Passwords should be secure, i.e. they should look random and should be hard to guess; they should be changed frequently, and should be different on different accounts of the same user; they should not be written down or stored in plain text.

The first idea for GPWs was described by Blonder (1996) \cite{Blonder-No-one-1996}. His approach was to let the user click, with a mouse or stylus, on a few chosen regions in an image that appeared on the screen. If the correct regions were clicked in, the user was authenticated, otherwise the user was rejected.

We think of giving users greater freedom, and greatly reflect personalized; exploring the relationship between facts and assumptions, there should be no gap between users and passwords.

``Graphical password'' will be abbreviated as GPW, and ``Topsnut-GPW'' is the abbreviation of ``Graphical passwords based on the idea of topological structure plus number theory''. We will use the knowledge of graph theory to design the so-called Topsnut-GPWs in \cite{bing-yao-others-2017-tianjin-university}.

GPW schemes have been proposed as a possible alternative to text-based schemes. Two principal research questions proposed by Wiedenbeck \emph{et al.} \cite{Wiedenbeck-Waters-Birget-Brodskiy-Memon-2005} are stated below:

\textbf{RQ 1:} Are GPWs a viable alternative to alphanumeric passwords in terms of security, as well as password creation, learning, performance, and retention?

\textbf{RQ 2:} Are users' perceptions of GPWs different from those of alphanumeric passwords?

How to answer the above problems? And furthermore can we meet more problems in researching GPWs?

Although people have designed many GPWs, no report tells us much of them were applied to business and practice (Ref. \cite{Suo-Zhu-Owen-2005}, \cite{Biddle-Chiasson-van-Oorschot-2009}, \cite{Gao-Jia-Ye-Ma-2013}). However, \emph{QR codes} (they are referred as a \emph{printable computer language}) can be considered as a type of GPWs that are used widely in the world, since it has the characteristics of large amount of information, strong error correcting ability, quick and comprehensive reading and so on. Clearly, QR code is a successful example of GPW' applications in mobile devices by fast, relatively reliable and other functions. Unfortunately, some of people can use QR codes to design traps for good people, since many people can not know their own QR codes and can not input their own QR codes by hand. The generation, recognition, scanning, making, decoding and designing of QR codes have become popular, in other words, QR codes can not be used to those places requiring high-level security.

By our observation, the existing GPWs wants people to learn more and remember more, and are lack of individualization, transformation, and persistent knowledge, etc. Many GPWs are not equal to users such that users have no right to choose their own formats or to make some of their favorite passwords. Needless to say, improving GPWs must be required.

\section{The existing GPWs}

The part of materials in this article are cited from three important surveys \cite{Suo-Zhu-Owen-2005}, \cite{Biddle-Chiasson-van-Oorschot-2009} and \cite{Gao-Jia-Ye-Ma-2013}.

\subsection{Basic constitutions of the existing GPWs}

\begin{asparaenum}[-]
\item Login Screen in public or private places: one picture/image or a group of pictures/images, or an $m\times n$ grid.

\item GPW's Length: number of pictures/images, number of click-points (stylus-points); number of drawing traces in grid.

\item  Order: order of pictures/images selected by users, order of positions in a series of click-points (resp. touch-points).

 \item  Personal replaceability: most of GPWs do not support personal replaceability frequently.

\item Geometric metric: geometric positions of click-points, touch-points, lines, curves in 2D-plane.

\item  Transformation: alphanumeric passwords can be transformed into pictures/images, and vice versa.

\item Round number of authentication: most of GPWs have only one round authentication.

\item  Compounding: Few number of GPWs consist of images and with alphanumeric passwords.

\item GPWs' spaces: many GPWs have small spaces.
\item  Pictures changed frequently: many of the existing GPWs' have no such function.
\end{asparaenum}

We point out that for being suitable to large number of people, most of the existing GPWs contain no mathematical computation. Also, it seems difficult to let most of people like a fixed picture/image when they input their graphical passwords.

\subsection{Possible attacks to the existing GPWs}

The existing attack types can be categorized as \emph{software attacks} and \emph{non-software attacks} including dictionary attack, shoulder surfing attack, hidden-camera spyware attack, social engineering attack, brute-force attack, intersection analysis attack, graphical dictionary attacks, guess attacks, smudge attacks, intersection analysis attack, and so on.

The most common of these attacks based on password space are common with the brute-force search and dictionary attack. Gao \emph{et al.} \cite{Gao-Jia-Ye-Ma-2013} have summarized the main attacks to the existing GPWs as follows:

\textbf{Shoulder surfing} refers to someone using direct observation techniques to capture passwords.

\textbf{Brute force attack} is also known as exhaustive-search attack, since it involves systematically searching all
possible elements in the theoretical password space until the correct one is found.

\textbf{Dictionary attack} involves guessing passwords from an exhaustive list called a dictionary (from a pre-arranged
list of values) which typically consisting of all passwords with higher possibility of being remembered easily, ordering from most to least probable.

\textbf{Intersection attack} is where all the password images are part of the challenge sets, and decoy icons are changed in each round. Intruders can use the intersection of two challenge sets to reveal the password images.

\textbf{Social Engineering} is a technique used by hackers or other attackers to gain access to seemingly secure systems through obtaining the needed information (for example, a username and password) from a person
rather than breaking into the system through electronic or algorithmic hacking techniques. Social Engineering has: (i) tricking; (ii) phishing and pharming.

\textbf{Spyware} is a type of malware (malicious software) installed on computers that collects information about
users without their knowledge. The presence spyware, which includes adware, Trojan horse, keystroke-loggers,
mouse-loggers and screen-scrapers, is usually installed on a user's personal computer without permission, is typically invisible to the user and difficult to detect. Spyware contains: (i) keystroke-loggers; (ii) mouse-loggers; (iii) screen-scrapers; (iv) other spyware.

In the end of the article \cite{Gao-Jia-Ye-Ma-2013}, the authors point out: ``(1) From a password scheme \emph{designer's perspective}, he must make his password scheme more secure and reliable, using methods where: Focus on increasing password entropy without sacrificing usability and memorability; minimize the pattern in the scheme; keyboard input or mouse click information not fixed for each login; add real-time SMS (Short Messaging Service) verification if necessary. (2) From a \emph{user's perspective}, he should make his password more secure by: Avoid pattern and easy password when set a password; use security antivirus software; not open unidentified web pages; not install suspicious plug-ins; not use websites requiring sensitive personal information in an insecure environment. However, for some systems which require high security levels, it is appropriate to sacrifice some usability to ensure the absolute security.''

\subsection{There are several rounds in the authentication of the existing GPWs}

The authors in \cite{Huanyu-Zhao-Xiaolin-Li-2007} introduce a so-called S3PAS for producing GPWs. This scheme seamlessly integrates both textual and GPWs and is resistant to shoulder-surfing, hidden-camera and spyware attacks. During the registration phase, users select a string $k=(k_1,k_2,\dots ,k_n)$ as the original password. The length of $k$ depends on different environments and different security requirements. During password creation, at the $i$th round, a user selects a letter/number $a_i$ in the $i$th triangle $\Delta k_{i}k_{i+1}k_{i+2}$ with $i=1,2,\dots,n$ by $n\equiv n$~$(\bmod~n)$ and $n+i\equiv i$~$(\bmod~n)$. After $n$ rounds, the user gets the desired password $a_1a_2\cdots a_n$.

\begin{rem} \label{rem:1111}
Topsnut-GPWs can be designed in many rounds in the process of authentication for meeting high security requirements.
\end{rem}

\subsection{GPWs for mobile devices}

GPWs for mobile devices have been investigated in \cite{Jansen-Gavrila-Korolev-Ayers-Swanstrom-2003}, \cite{Takada-Koike-2003},
\cite{Dunphy-Heiner-Asokan-2010}, \cite{Xiaoyuan-Suo-2014} and \cite{YAO-SUN-ZHANG-LI-YAN-2017}. Suo, in her article \cite{Xiaoyuan-Suo-2014}, has proposed the following suggestions:
\begin{asparaenum}[-]
\item Approaches to overcome limitations of a touch screen computer for graphical
password designs.

\item The relationship between user password choices and the complexity of the
background image.

\item The relationship between background image choice and successful authentication
rate.

\item The relationship between tolerance rate and successful rate.

\item Security concerns of using GPW for touch screen devices.

\item Assess the future of GPW for touch screen devices.
\end{asparaenum}

\begin{figure}[h]
\centering
\includegraphics[height=8cm]{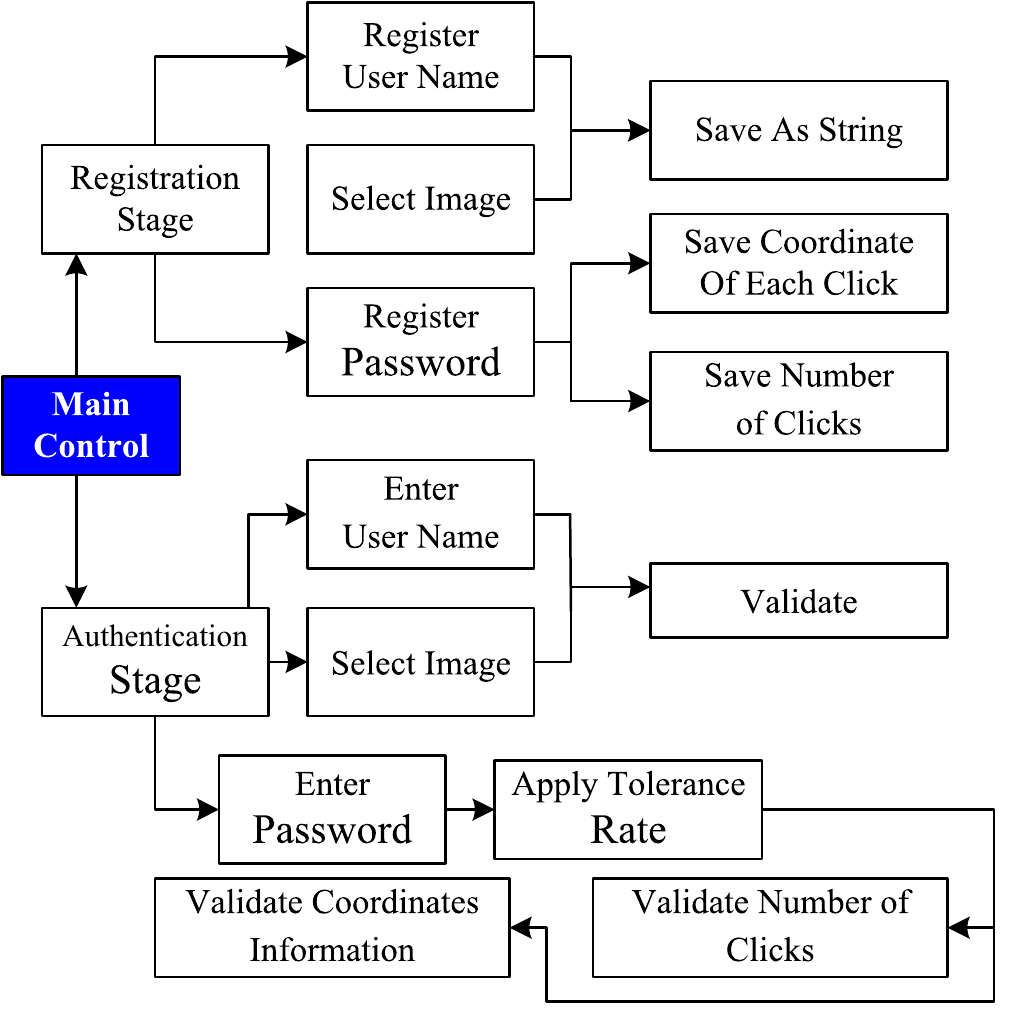}\\
\caption{\label{fig:suo-tu}{\footnotesize The process of registration and authentication \cite{Xiaoyuan-Suo-2014}.}}
\end{figure}

Suo shows her GPW for mobile devices (see Fig. \ref{fig:suo-tu}).

\section{Topsnut-GPWs for mobile devices}

\begin{figure}[h]
\centering
\includegraphics[height=3.5cm]{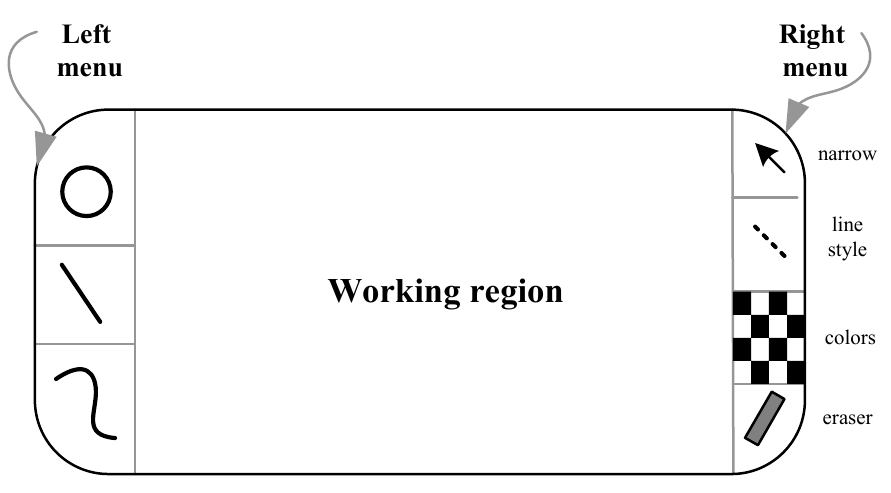}\\
\caption{\label{fig:mobile-00}{\footnotesize The screen for Topsnut-GPWs.}}
\end{figure}

\begin{figure}[h]
\centering
\includegraphics[height=4.2cm]{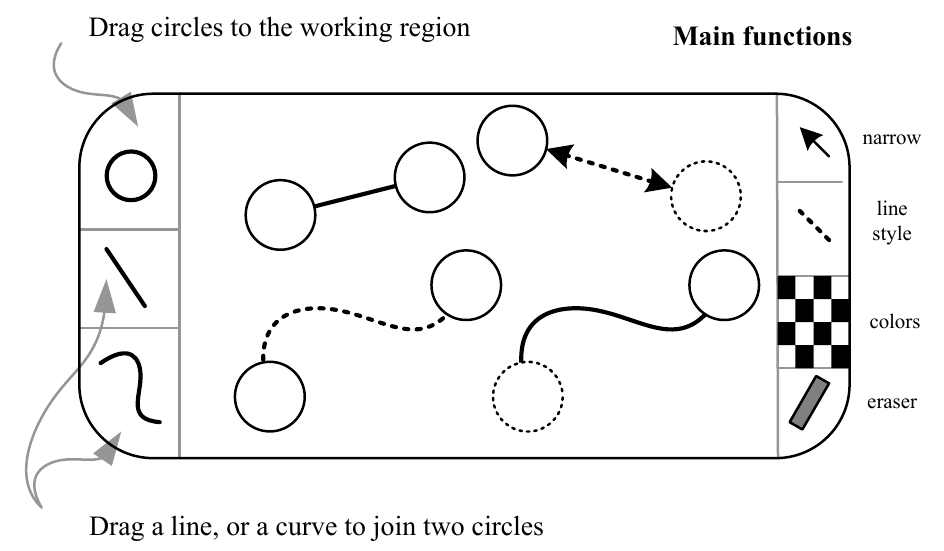}\\
\caption{\label{fig:mobile-0}{\footnotesize The beginning screen of the Topsnut-GPW' software on a mobile. }}
\end{figure}

\begin{figure}[h]
\centering
\includegraphics[height=3.8cm]{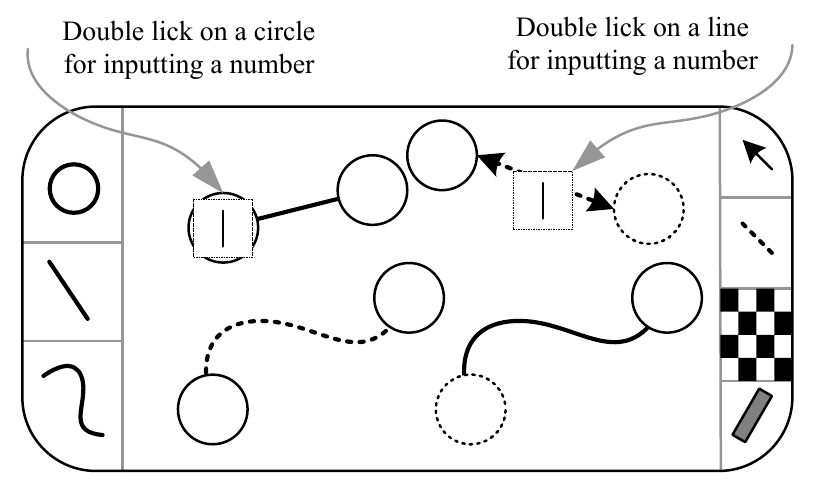}\\
\caption{\label{fig:mobile-1}{\footnotesize Double lick to the center of any circle
for inputting a number/letter, and double lick to the center of any line/curve
for labeling with a number/letter.}}
\end{figure}

\begin{figure}[h]
\centering
\includegraphics[height=3cm]{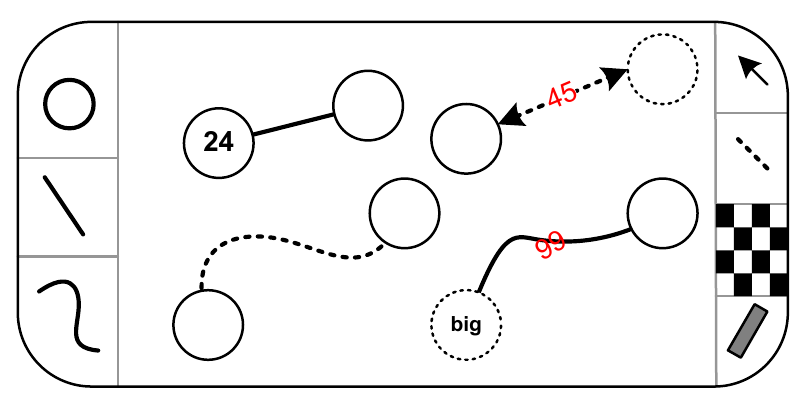}\\
\caption{\label{fig:mobile-2}{\footnotesize Some circles and lines are labeled with numbers or letters.}}
\end{figure}

\begin{figure}[h]
\centering
\includegraphics[height=3cm]{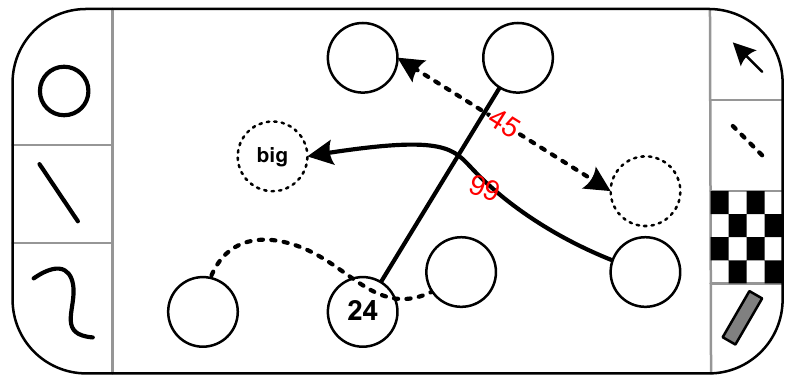}\\
\caption{\label{fig:mobile-3}{\footnotesize Drag a circle to any place in the working region, the line/curve joining this circle with another circle will follow this circle, like the case in Microsoft Office Visio; and add narrows to some lines or curves.}}
\end{figure}

We use standard notation and terminology of graph theory that can be found in \cite{Bondy-2008}, in which there are many graph colorings were introduced and investigated by algorithmic methods. Gallian \cite{Gallian2016} presents a large survey on graph labellings, over 2000 papers collected. We present our investigation of GPWs on mobile devices by using Topsnut-GPWs based on the idea appeared in \cite{Wang-Xu-Yao-2016} and \cite{Wang-Xu-Yao-Key-models-Lock-models-2016}. In Fig.\ref{fig:mobile-00}, we show a screen for Topsnut-GPWs, in which there are three regions, we drag small circles from the left menu to the working region, and then drag a line or a curve to join some pairs of small circles in the working region. In the above procedure, small circles and lines (or curves) can automatically past to each other like that in Microsoft office Visio (see Fig. \ref{fig:mobile-0}). The following process is to label the circles and lines with numbers or letters (see Fig. \ref{fig:mobile-1}-Fig. \ref{fig:mobile-3}).

\subsection{Topsnut-GPWs}

In two articles \cite{Wang-Xu-Yao-2016} and \cite{Wang-Xu-Yao-Key-models-Lock-models-2016}, Wang \emph{et al.} show an idea of ``topological structures plus number theory'' for designing new-type GPWs (abbreviated as Topsnut-GPWs), and have designed some Topsnut-GPWs in the techniques of graph theory. Wang \emph{et al.} \cite{Wang-Xu-Yao-2016} are interesting on devising Topsnut-GPWs for mobile devices with touch screen, such as smart phones, iPad and those are popular hand-held touch devices. By their principle of ``needing conditions as little as possible, maximizing users' needs as large as possible'' they develop Topsnut-GPWs for different accounts, easily remembering, frequently changing, embodying, individual favorite, withstanding popular attacks. Clearly, the key on generating Topsnut-GPWs is to increase usability and security simultaneously (Ref. \cite{Biddle-Chiasson-van-Oorschot-2009}).

\begin{defn} \label{defn:Topsnut-GPW}
 Let $G$ be a graph of graph theory, and $f$ be a coloring/labelling defined as $f:X\rightarrow M$, where $X$ is a subset of $V(G)\cup E(G)$, and $M$ is an integer set. We call this labeled graph $G$ a \emph{Topsnut-GPW}.
\end{defn}

\begin{figure}[h]
\centering
\includegraphics[height=2.6cm]{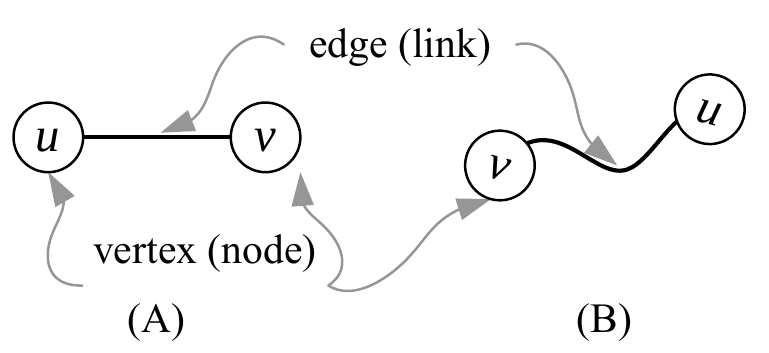}\\
\caption{\label{fig:graph-1}{\footnotesize An \emph{edge} $uv$ has two ends $u$ and $v$ that are called \emph{vertices} in graph theory. The edge $uv$ can be expressed by a line or by a curve, so (A) and (B) are the same by the view of graph theory. }}
\end{figure}

\begin{figure}[h]
\centering
\includegraphics[height=2cm]{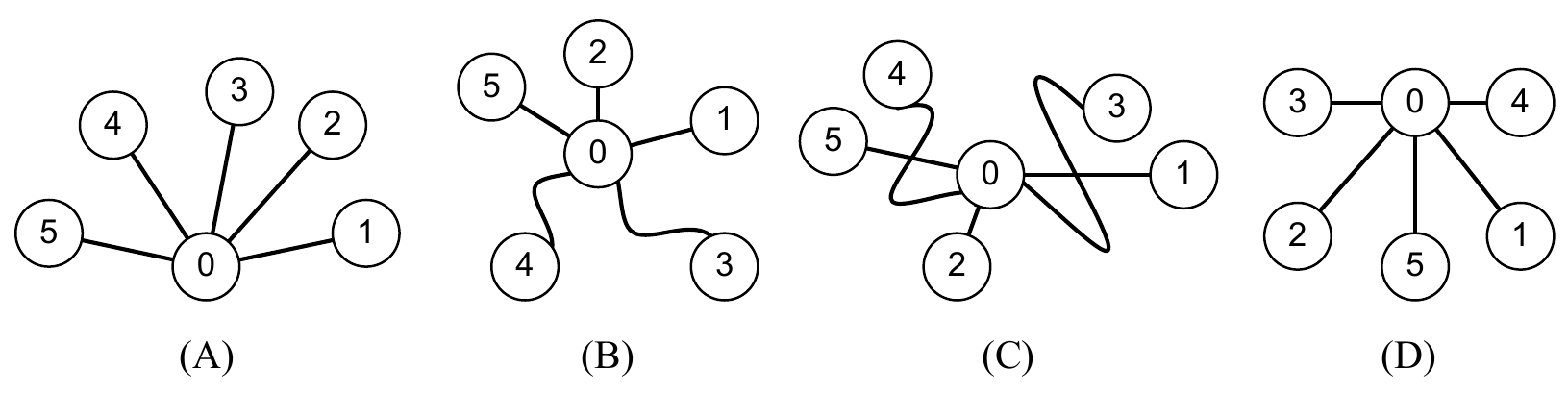}\\
\caption{\label{fig:graph-2}{\footnotesize Four labeled graphs (A), (B), (C) and (D) are one Topsnut-GPW indeed. }}
\end{figure}

\begin{figure}[h]
\centering
\includegraphics[height=2cm]{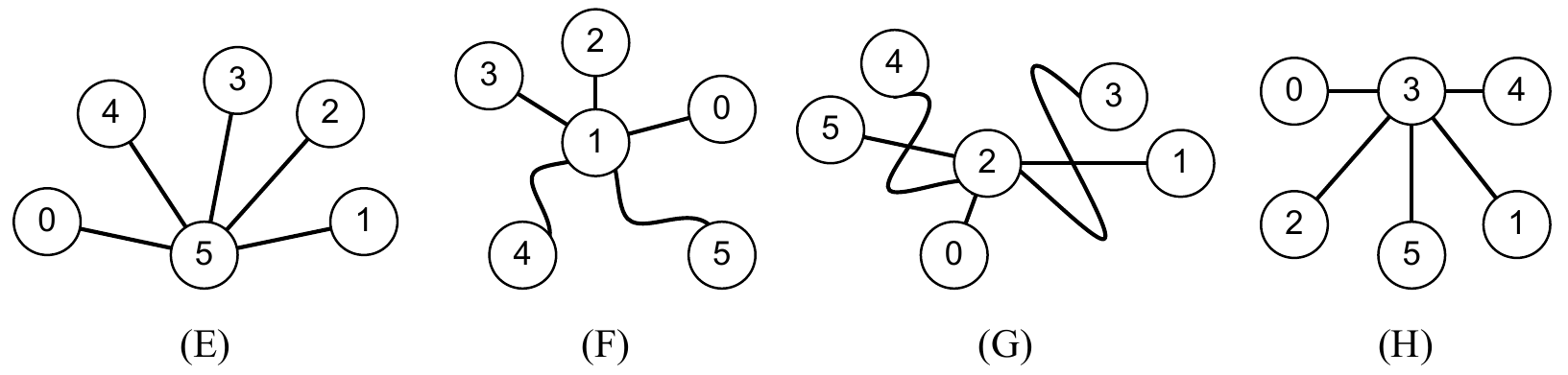}\\
\caption{\label{fig:graph-3}{\footnotesize Four labeled graphs (E), (F), (G) and (H) are different to each other under the meaning of Topsnut-GPWs, and they differ from any one of four labeled graphs shown in Fig.\ref{fig:graph-2}. }}
\end{figure}

Fig.\ref{fig:graph-1} shows an edge $uv$ that joins two vertices $u$ and $v$, also, we call two vertices $u$ and $v$ as the ends of the edge $uv$. In graph theory, two vertices $u$ and $v$ can be joined by a line or a curve to form an edge $uv$, and they are drawn in 2D-plane by no requirements of geometric metrics. There are four Topsnut-GPWs shown in Fig.\ref{fig:graph-2}, in fact, they are one Topsnut-GPW by the definition of Topsnut-GPWs. The development goals of Topsnut-GPWs are stated in the following:

1. Both users and authentication need little storage space.

2. User needs little scientific knowledge of mathematics, and authentication can identify more complex scientific knowledge of mathematics, chemistry, physics and biology.

3. Operate by human fingers on mobile devices with touch screen, which can be used at home or private places for resisting shoulder surfing attack or other physical attacks.

4. Only need small circles, line/curve segments (colored, colorless, continuous and dotted lines/curves); small circles and line segments can be dynamically connected up together, and latters/numerals can be marked on small circles and line/curve segments.

5. Part of Topsnut-GPWs can be privately customization, that is, let users construct and select their favorite and non-forgetting topological structures, as well as choose their own mathematical or non-mathematical techniques.

6. To achieve successfully transformation between low-level passwords and advanced passwords.

7. Pursuit of simple yet quick principles.

8. Inherit the advantages of traditional graphical cryptography and two-dimensional codes (QR code) as much as possible.

9. Take into account the emergence of \emph{intelligent graphical passwords} (smart GPWs) for achieving sustainable development.

\subsection{Examples of Topsnut-GPWs}

\begin{figure}[h]
\centering
\includegraphics[height=7.4cm]{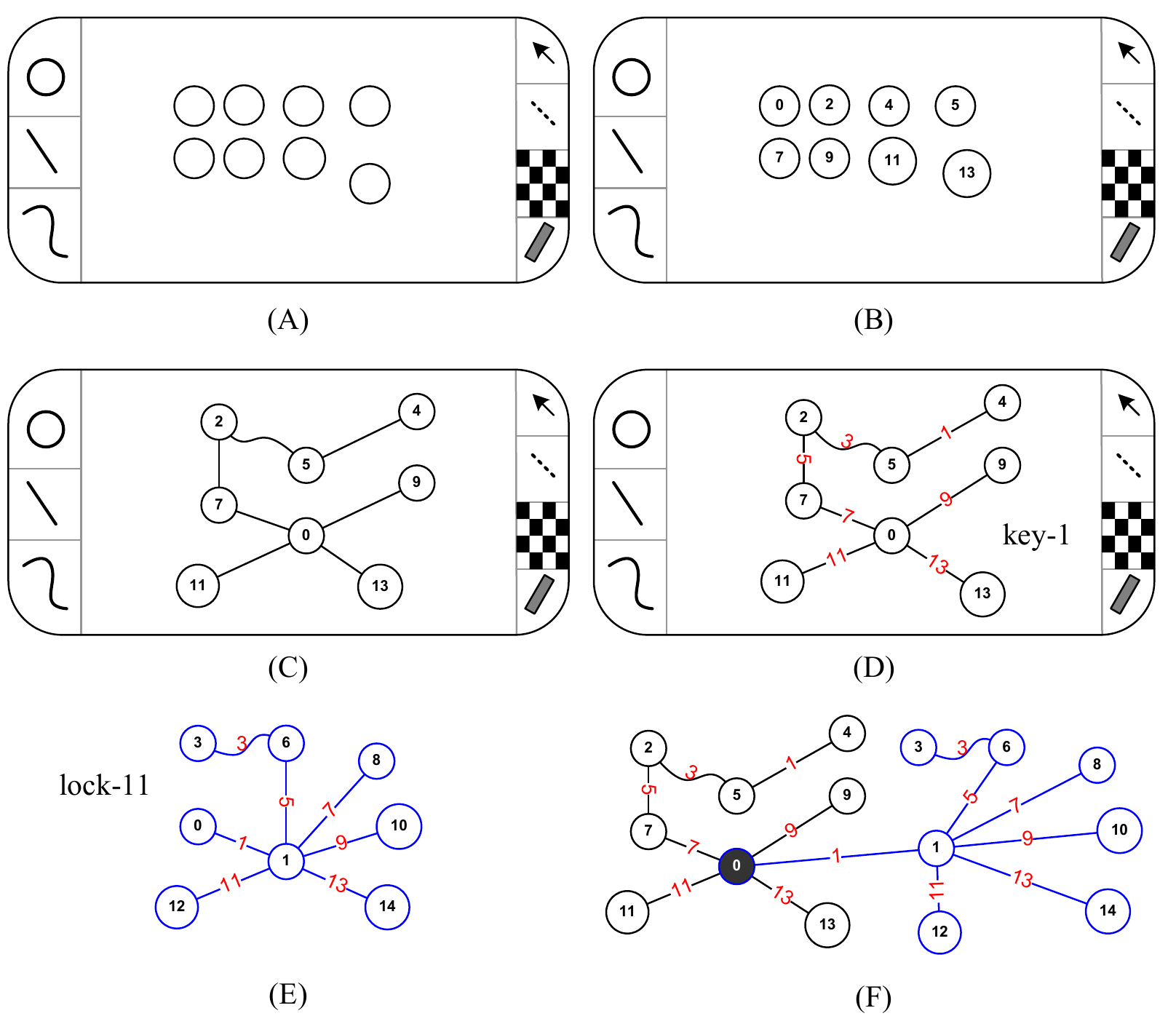}\\
\caption{\label{fig:GPW-1}{\footnotesize A procedure of making a simple Topsnut-GPW: (A) Drag small circles into the working region; (B) mark the small circles with numbers; (C) join some pairs of labeled small circles by lines or curves by your like; (D) label the lines or curves with numbers, the key is made well, and send it for authentication; (E) the key and the lock form a complete authentication. }}
\end{figure}

First example is shown in Fig.\ref{fig:GPW-1}, we can see how to generate a simple Topsnut-GPW.

A user is logging for his business, he got the first screen shown in Fig.\ref{fig:GPW-1}(A) after inputting his account. He dragged several small circles into the working region, see Fig.\ref{fig:GPW-1}(B), and labeled the small circles with numbers shown in Fig.\ref{fig:GPW-1}(C). The authentication shown in Fig.\ref{fig:GPW-1}(F) is a combined Topsnut-GPW $G$ made by a key Fig.\ref{fig:GPW-1}(D) and a lock Fig.\ref{fig:GPW-1}(E). This Topsnut-GPW $G$ was designed first in \cite{Wang-Xu-Yao-2017}, called a \emph{twin odd-graceful labelling}, and it has its own vertex labels $0,1,2,\dots, 14$ and its own edge labels form two odd-integer sets $\{1,3,5, 7,9,11,13\}$ and $\{1,3,5, 7,9,11,13\}$. It is easy to see that the number labeled to an edge just equals the absolute value of difference of two numbers labeled to two small circles that are the ends of this edge.

In Fig.\ref{fig:GPW-1}, three steps (B), (C) and (D) can be arranged into other orders. For example, we can join some pairs of unlabeled small circles for making a topological structure selected by users, and then label the small circles (or lines or curves) with numbers. Different orders are very important, since they will meet the records in authentication.

\begin{figure}[h]
\centering
\includegraphics[height=4.8cm]{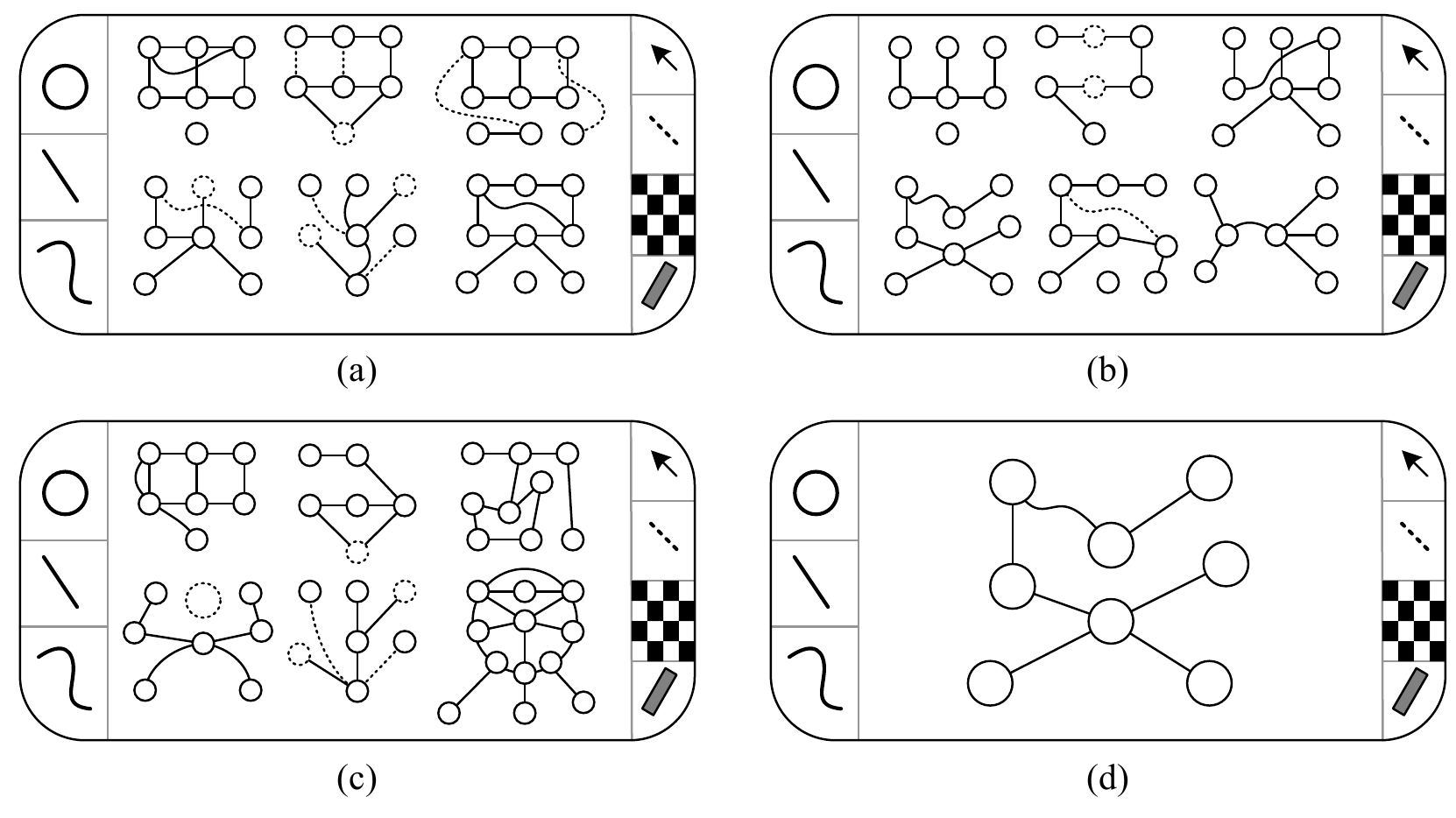}\\
\caption{\label{fig:GPW-2}{\footnotesize Three screens (a), (b) and (c) will appear circularly before the user select one topological structure; (d) is the selected topological structure for making a key. }}
\end{figure}

Second example is shown in Fig.\ref{fig:GPW-2}. At the first step of log in, the user get three screens (a), (b) and (c), which will appear circularly before the user select one topological structure. The selected topological structure shown in Fig.\ref{fig:GPW-2}(d) is as the same as one shown in Fig.\ref{fig:GPW-1}(D). Next the user will label the small circles and lines of the topological structure with numbers such that three numbers labeled to an edge and two ends of the edge meet a predetermined requirement.

\subsection{Properties of Topsnut-GPWs}

All of human activities can be expressed in ``language'' we have thought, where ``language'' is a combination of pictures, sounds, videos, scientific languages, scientific symbols, scientific knowledge, biological techniques, etc. So, Topsnut-GPW is a particular ``language'', and can be considered as a \emph{platform} for designing various Topsnut-GPWs.

\vskip 0.2cm

\subsubsection{Properties} The study of the ``language'' has the significance of graph theory and practical application. We have the following advantages of ``language'':
\begin{asparaenum}[C-1.]
\item There is a vast number of graphs with smaller orders (see \cite{Harary-Palmer-1973}).

\item There exist enormous numbers of graph colorings and labellings in graph theory (see \cite{Gallian2016}). And new graph colorings/labellings come into being everyday.

\item For easy memory, some simpler operations like addition, subtraction, absolution and finite modular operations are applied.

\item There are many non-polynomial algorithms. For example, drawing non-isomorphic graphs is very difficult and non-polynomial; for a given graph, finding out all possible colorings/labellings are impossible, since these colorings/labellings are massive data. Many graph problems are NP-complete.

\item As known, tree structures can adapt to a large number of labellings, in addition to graceful labeling, no other labellings reported that were established in those tree structures having smaller vertex numbers by computer, almost no computer proof. Because construction methods are complex, this means that using computers to break down GPWs will be difficult greatly.

\item The number of one style of different labellings/colorings is large, and no method is reported to find out all of such labellings/colorings.

\item There are many mathematical conjectures (open problems) in graph labellings, such as the famous graceful tree conjecture, odd-graceful tree conjecture, etc.

\item Many labellings of trees are convertible to each other (see \cite{Yao-Liu-Yao-2017}).

\item Topsnut-GPWs are suitable to a wide range of people, since they have much interesting, strongly mathematical logic and many mathematical conjectures.

\item One key corresponds to more locks, or more keys corresponds to one lock only.

\item Topsnut-GPWs realize the coexistence of two or more labellings on a graph, which leads to the problem of multi-labelling decomposition of graphs, and brings new research objects and new problems to graph theory.

\item There are connections between Topsnut-GPWs and other type of passwords. For example, small circles in the Topsnut-GPWs can be equipped with fingerprints and other biological information requirements, and users' pictures can be embedded in small circles, greatly reflects personalization.

\item Number theory, algebra and graph theory are the strong support to Topsnut-GPWs.

\item Topsnut-GPWs are easy to avoid non-software attacks, since people can use their mobile devices in private or semi-private places. However, Topsnut-GPWs should resist various spyware like Trojan horse viruses.
\end{asparaenum}

\vskip 0.2cm

\subsubsection{Connections of Topsnut-GPWs} By means of graph labellings, Wang \emph{et al.} \cite{Wang-Xu-Yao-2017} show some new graph labellings in the procedure of building Topsnut-GPWs such that a graph can admits two different labellings. Based on trees, Yao \emph{et al.} \cite{Yao-Liu-Yao-2017} show the equivalent connections among eight different labellings under the set-ordered graceful condition.

\subsection{Construction of Topsnut-GPWs}

\begin{figure}[h]
\centering
\includegraphics[height=9cm]{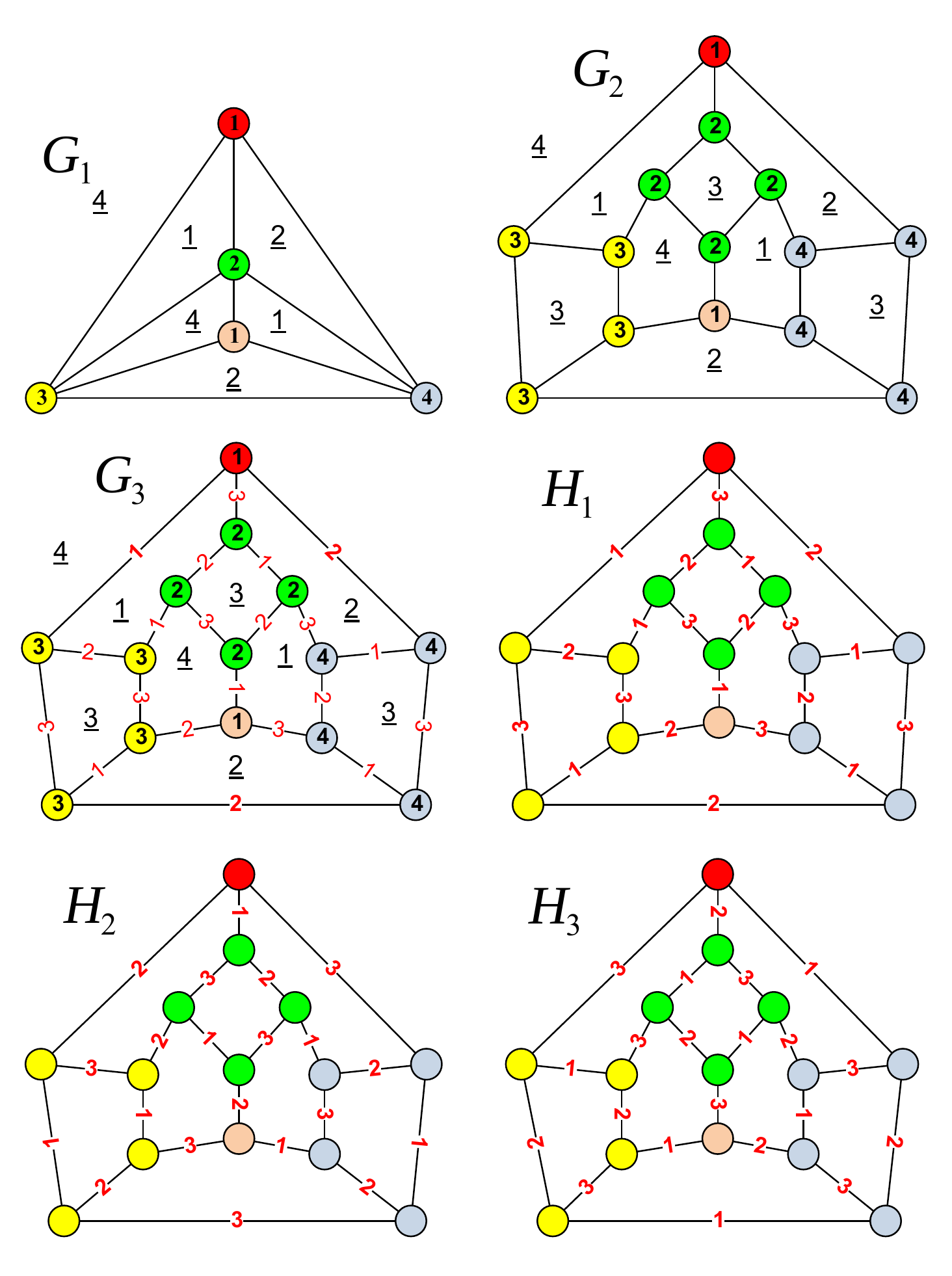}\\
\caption{\label{fig:xu-00}{\footnotesize Some connection between colorings of planar graphs. }}
\end{figure}

We can apply many constructive techniques and graph colorings/labellings of graph theory on the \emph{Topsnut platform} when we design a myriad of Topsnut-GPWs. An example shown in Fig. \ref{fig:xu-00} introduce the connection between colorings of planar graphs. In Fig. \ref{fig:xu-00}, $G_1$ is a maximal planar graph having a 4-coloring and a 3-face-coloring; $G_2$ is a 3-regular planar graph obtained from $G_1$; the \emph{Klein four-group} enables us to obtain $G_3$ having a proper 3-edge-coloring. The deletion of numbers of vertices and faces yields $H_1$, and add one to the edge labels of $H_1$ under modular $3$, where $3\equiv 3$ and $0\equiv 3~(\bmod~3)$, to get $H_2$. By the same way in producing $H_2$ we can obtain $H_3$. It is not hard to see that $f_i(uv)+f_j(uv)-f_k(uv)=f_{i+j-k~(\bmod~3)}(u)$ with $0\equiv 3~(\bmod~3)$, where $f_l(uv)$ is the label of edge $uv$ of $H_l$ with $l=1,2,3$. So, three Topsnut-GPWs $H_1$, $H_2$ and $H_3$ form an \emph{Abelian additive graphical group of modular 3}, and each $H_i$ can be considered as the unit element of this graphical group.

\begin{rem} \label{rem:22222}
Such phenomenon are studied in \cite{Yao-Sun-Zhao-Li-Yan-2017} and \cite{Sun-Zhang-ZHAO-Yao-2017}.
\end{rem}

\subsubsection{Recursive graphs} We define a class of \emph{recursive graphs} $G_n~(n\geq 0)$ in the way:

\begin{defn} \label{defn:recursive-graphs}
Let $O_r$ be an operation defined on graphs. If each $G_{n}$ can be produced from $G_{n-1}$ by the operation $O_r$, and $G_0$ is not generated by doing the operation $O_r$ to some graph, then we call $G_{n}$ a \emph{recursive $O_r$-graph} and $G_0$ a \emph{$O_r$-root}, as well as $O_r$ a \emph{recursive operation}.
\end{defn}

The recursive operations contain the \emph{triangularly edge-identifying operation}, the \emph{triangularly embedded edge-overlapping operation} and the \emph{triangularly single-edge-paste operation}, and so on.

\begin{figure}[h]
\centering
\includegraphics[height=2.6cm]{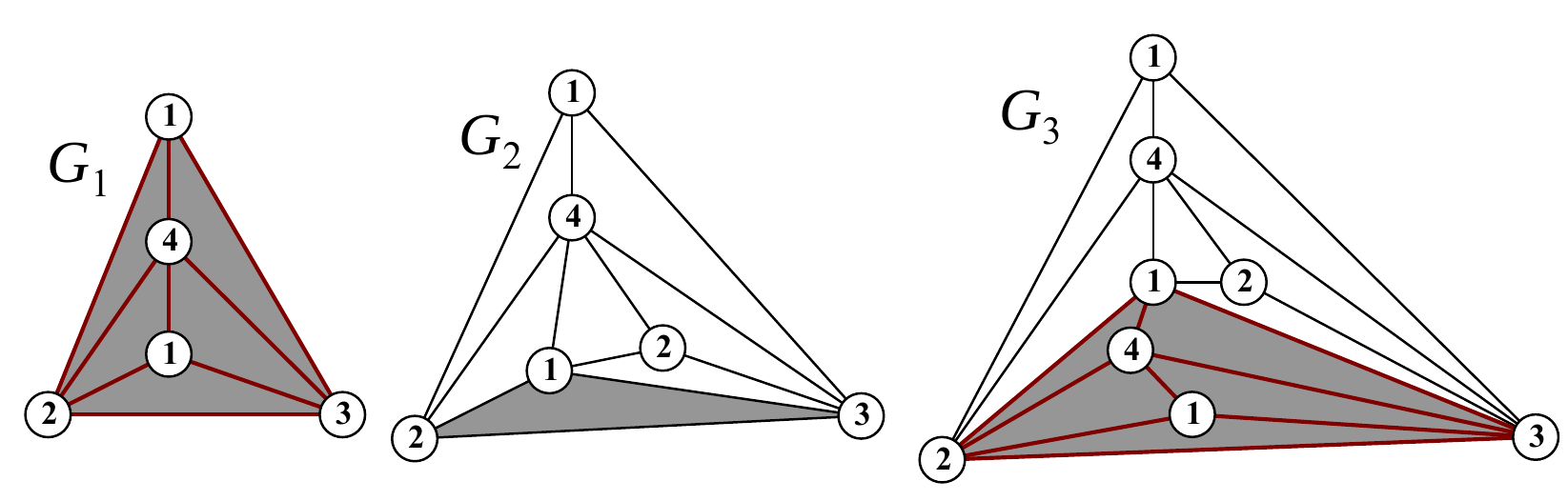}\\
\caption{\label{fig:Triangular-operation-00}{\footnotesize A scheme of a triangularly embedded edge-overlapping operation. }}
\end{figure}

A \emph{triangularly embedded edge-overlapping operation} (TEEoO) on maximal planar graphs is defined as: Let $G$, $H$ be two maximal planar graphs, where $G$ has its own inner face bound $\Delta ABC$ and $H$ has its own outer face bound $\Delta abc$. We embed $H$ into $G$ such that the edge $ab$ of the outer face bound of $H$ is overlapped with the edge $AB$ of the inner face bound of $G$ into one, and do the same operation on the edges $bc$ and $BC$, on the edges $ca$ and $CA$. The resulting graph is called a \emph{TEEoO-graph}, denoted as $G\Delta (H)$, and call $H$ a \emph{TEEoO-factor} and $G$ a \emph{TEEoO-object}. Clearly, any recursive maximal planar graph $G'$ is a TEEoO-graph obtained by do a TEEoO to a inner face of a recursive maximal planar graph $G$ with the TEEoO-factor $K_4$, namely, $G'=G\Delta (K_4)$. In fact, any recursive maximal planar graph $G'$ has the $O_r$-root $K_3$, where $O_r=$TEEoO (see an example shown in Fig. \ref{fig:Triangular-operation-00}). Other two examples are Apollonian network model and recursive maximal planar graphs.

\begin{figure}[h]
\centering
\includegraphics[height=2.6cm]{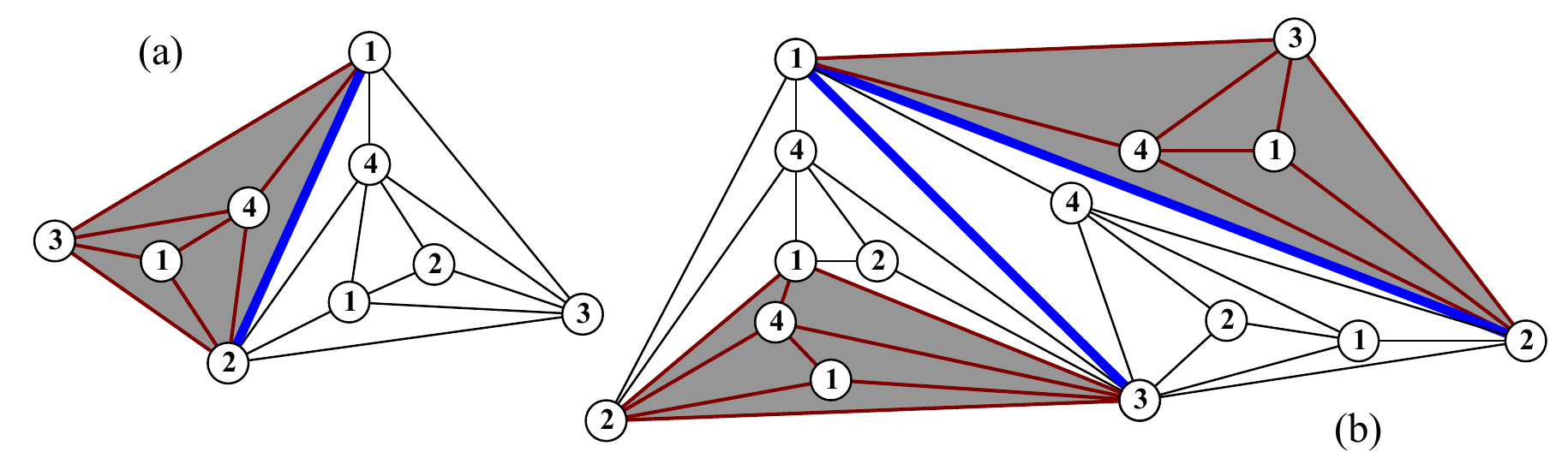}\\
\caption{\label{fig:Triangular-operation-11}{\footnotesize (a) is obtained from $G_1$ and $G_2$ in Fig. \ref{fig:Triangular-operation-00} by a triangularly single-edge-paste operation on an edge (thick); (b) is obtained from (a) and $G_3$ in Fig. \ref{fig:Triangular-operation-00} by a triangularly single-edge-paste operation on an edge (thick).}}
\end{figure}

\begin{rem} \label{rem:33333}
If a triangularly embedded edge-overlapping operation (TEEoO) based on maximal planar graphs, that is, the TEEoO-factor and the TEEoO-object are maximal planar graphs, then we say TEEoO will close many properties of the TEEoO-factor and the TEEoO-object, such as various 4-colorings. The same situations appear in the graphs made by the triangularly single-edge-paste operation. Sierp\'{i}nski network model is constructed by another type of triangularly recursive operation.
\end{rem}

\vskip 0.2cm

\textbf{Problem:} (i) List recursive operations $O_r$ often used, and recursive $O_r$-graphs appeared in networks.

(ii) For a recursive operation $O_r$, we show the characters of the recursive $O_r$-graphs $G_{n}$ and determine the $O_r$-root $G_0$.

\vskip 0.2cm

\subsubsection{Xu's methods based on maximal planar graphs}

First of all, in designing Topsnut-GPWs, we introduce useful and powerful Xu's methods in his articles \cite{Jin-Xu-(1)-2016, Jin-Xu-(2)-2016,Jin-Xu-(3)-2016,Jin-Xu-(4)-2016} on his mathematical proof of Four Color Conjecture. The reasons of establishing Topsnut-GPWs by using Xu's methods are:
\begin{asparaenum}[(1)]
\item Easy to remember: use only 4 numbers (alternatively, 4 colors, or 4 letters, or 4 pictures, or 4 types of circles, etc.)

\item The mathematical theory guarantees that every planar graph has a proper 4-coloring (\cite{Jin-Xu-(1)-2016, Jin-Xu-(2)-2016,Jin-Xu-(3)-2016,Jin-Xu-(4)-2016}).
\item Maximal planar graphs have: (1) Normative standard, each face of any maximal planar graph is a triangle; (2) configuration complexity. Using computer to construct maximal planar graphs is generally irregular, and determining the number of non-isomorphic maximal planar graphs is NP-hard.
\item It is difficult to find all 4-colorings of a maximal planar graph. General attackers are not able to find 4-colorings of maximal planar graphs by computer. Nearly 50 years, only two reports by American scientists overcome the planar graph 4-coloring problem by computer and long working time.
\item Interchangeability. Topsnut-GPWs can be contractible and extensible; the operations of graph theory can implement conversions between low-level cryptography and high-level cryptography.
\end{asparaenum}

We show methods on maximal planar graphs for designing Topsnut-GPWs' structures as follows.

\vskip 0.2cm

\begin{figure}[h]
\centering
\includegraphics[height=3.2cm]{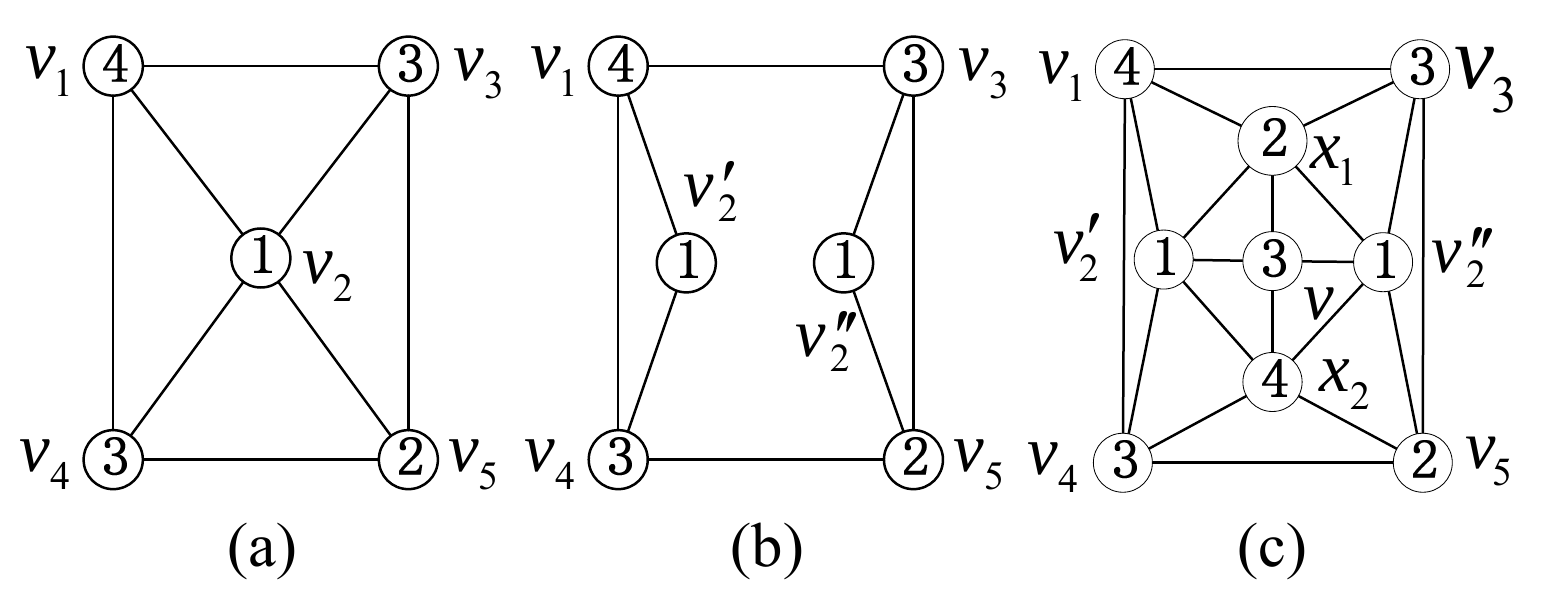}\\
\caption{\label{fig:dumbbell-11}{\footnotesize A dumbbell transformation 1. }}
\end{figure}

\textbf{Method 1.} The dumbbell transformation (\cite{Jin-Xu-(2)-2016, Jin-Xu-(3)-2016}). Fig.\ref{fig:dumbbell-11} (a) is called a \emph{dumbbell graph} in Xu's operation. We cut the vertex $v_2$ shown in Fig.\ref{fig:dumbbell-11}(a) into two subvertices $v'_2,v''_2$ for getting Fig.\ref{fig:dumbbell-11} (b), and Fig.\ref{fig:dumbbell-11} (c) is obtained by doing a dumbbell transformation to Fig.\ref{fig:dumbbell-11} (b).

In Fig.\ref{fig:dumbbell-11} (c), from a large cycle 3'4323' to a small cycle 1'2141' , and from the small cycle 1'2141' to the center vertex 3, we can get an alphanumeric password 3'4323'1'2141'3'3', it has 18 units. Moreover, based on a path $4141=v_1v'_2x_2v''_2$ in Fig.\ref{fig:dumbbell-11} (c), we can get another alphanumeric password $PW_1=(v_1=)4'33214'(v'_2=)1'342341'(x_2=)4'313124'(v''_2=)1'432321' $ having 35 units.

\begin{figure}[h]
\centering
\includegraphics[height=3.2cm]{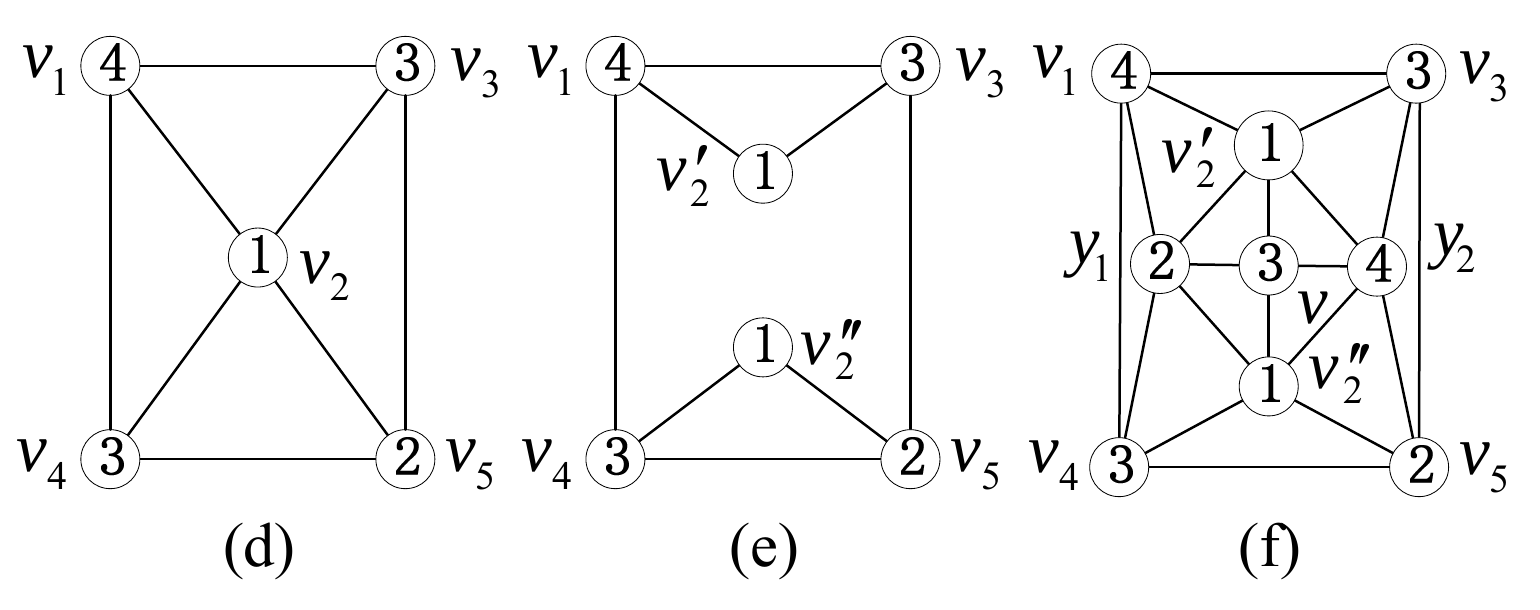}\\
\caption{\label{fig:dumbbell-22}{\footnotesize A dumbbell transformation 2. }}
\end{figure}

Moreover, in Fig.\ref{fig:dumbbell-22} (f), the path $4141=v_1v'_2y_2v''_2$ can yields an alphanumeric password $PW_2=(v_1=)4'343124'(v'_2=)1'324341'(y_2=)4'313214'(v''_2=)1'323421'$ with 35 units. Clearly, $PW_1\neq PW_2$, although Fig.\ref{fig:dumbbell-11} (a) coincides with Fig.\ref{fig:dumbbell-22} (d). Moreover, in Fig.\ref{fig:dumbbell-11} (c), we can use a path $32323=v_4v_5v_3x_1v$ to get an alphanumeric password $PW_3=(v_4=)3'41423'(v_5=)2'34132'(v_3=)3'21243'(x_1=)2'131432'(v=)3'14123'$ with 41 units, and another path $32323=vy_1v_4v_5v_3$ in Fig.\ref{fig:dumbbell-22} (f) to get another alphanumeric password $PW_4$ such that $PW_3\neq PW_4$. Or, we can use a cycle $C_1=v_1v_3v_5v_4x_2v''_2x_1vv'_2v_1$ to make a long-unit alphanumeric password $PW_5$ such that $PW_4\neq PW_5$.

\begin{rem} \label{rem:44444}
(1) Each of labeled graphs in Fig.\ref{fig:dumbbell-11} and Fig.\ref{fig:dumbbell-22} is just a Topsnut-GPW.

(2) It is obviously difficult to draw completely Fig.\ref{fig:dumbbell-11} (c) and Fig.\ref{fig:dumbbell-22} (f) by two alphanumeric passwords $PW_1$ and $PW_2$.

(3) Since there are many Xu's dumbbell graphs in Fig.\ref{fig:dumbbell-11} (c) and Fig.\ref{fig:dumbbell-22} (f), we can implement the dumbbell transformation on these two Topsnut-GPWs for generating more complex Topsnut-GPWs and producing alphanumeric passwords having enough long units (they can be used to encrypt electronic documents, or to produce encryption keys) from these more complex Topsnut-GPWs.
\end{rem}

\begin{figure}[h]
\centering
\includegraphics[height=6cm]{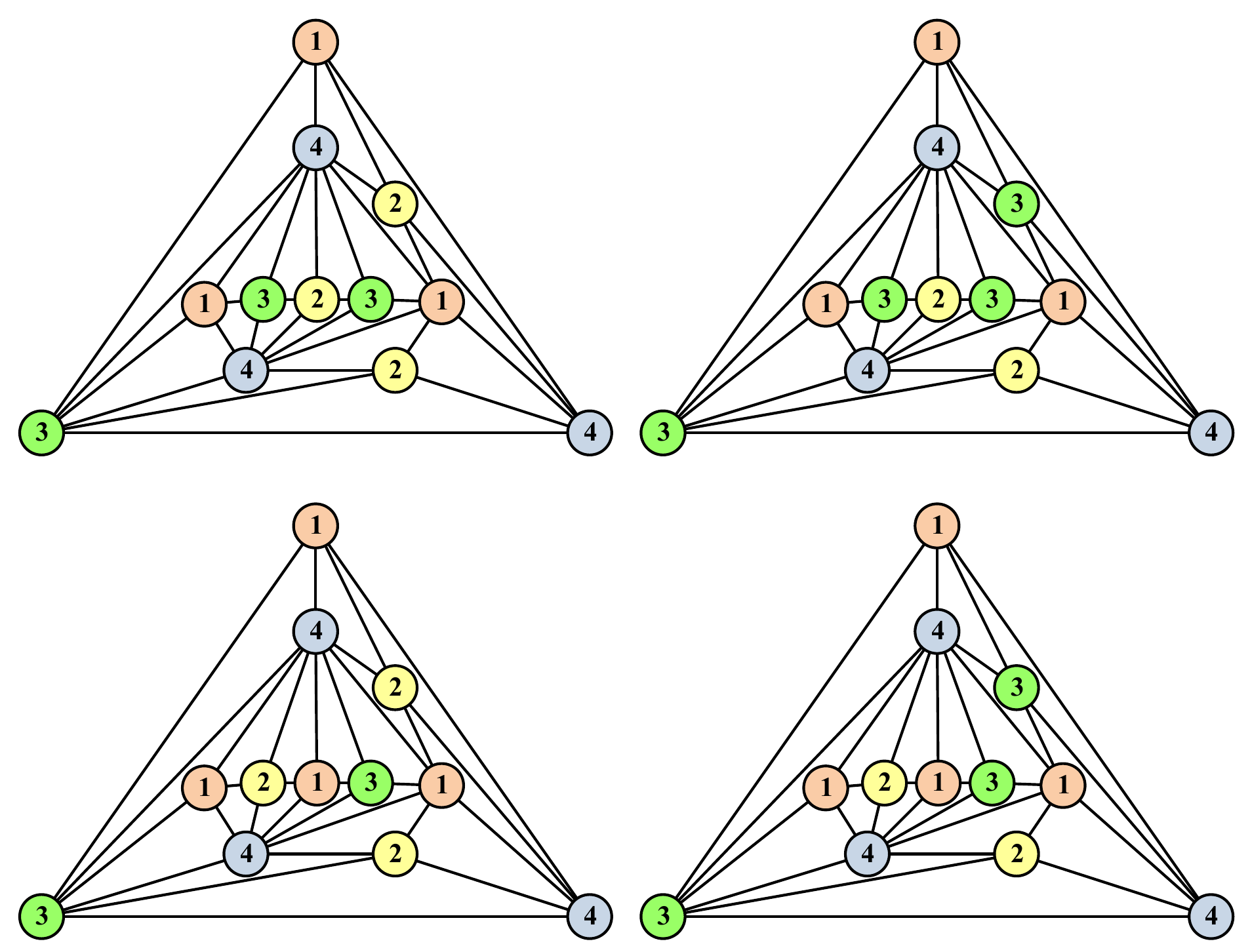}\\
\caption{\label{fig:dumbbell-22}{\footnotesize Four Topsnut-GPWs obtained by Xu's transformation. }}
\end{figure}

\vskip 0.4cm

\textbf{Method 2.} How many Topsnut-GPWs are there based on a maximal planar graph? In the view of authentication, this problem is very important such that two users have different Topsnut-GPWs if they have selected just the same maximal planar graph. A \emph{Kempe change} is to exchange two colors of a connected component of a 2-coloring induced subgraph, and remain the colors of the other vertices unchanged in a colored graph. Two $k$-colorings $f$ and $f'$ of a $k~(\geq 2)$-chromatic graph $G$ are called \emph{Kempe equivalent} if $f'$ can be obtained from $f$ by a sequence of \emph{Kempe changes}. Xu, in his article \cite{Jin-Xu-(4)-2016}, discover the Kempe's equivalent class: Let $G$ be a $k$-chromatic graph. $G$ is called a \emph{Kempe graph} if all $k$-colorings of $G$ are \emph{Kempe equivalent}. And he studies the characteristics of \emph{Kempe maximal planar graphs}, introduce the \emph{recursive domino method} to construct Kempe maximal planar graphs, and propose two interesting conjectures. Xu partitions the Kempe equivalent classes of non-Kempe graphs into three classes: \emph{tree-type}, \emph{cycle-type}, and \emph{circular-cycle-type}, and point out that all these three classes can exist simultaneously in the set of 4-colorings of one maximal planar graph \cite{Jin-Xu-(4)-2016}.

\vskip 0.4cm

\begin{figure}[h]
\centering
\includegraphics[height=4.6cm]{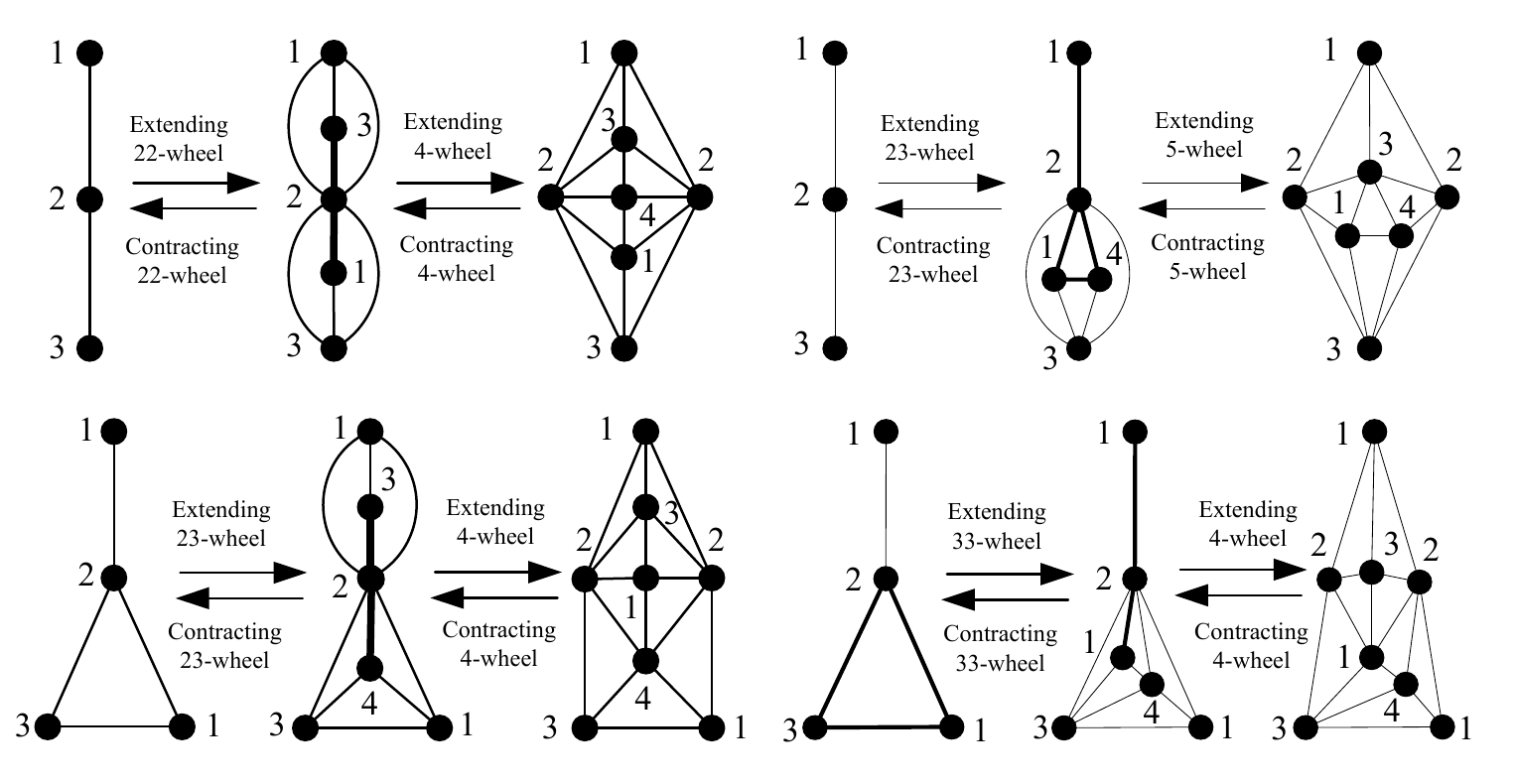}\\
\caption{\label{fig:Domino-1}{\footnotesize Four domino configurations with three vertices as the centers of 3-wheel, 4-wheel and 5-wheel. }}
\end{figure}

\begin{figure}[h]
\centering
\includegraphics[height=4.2cm]{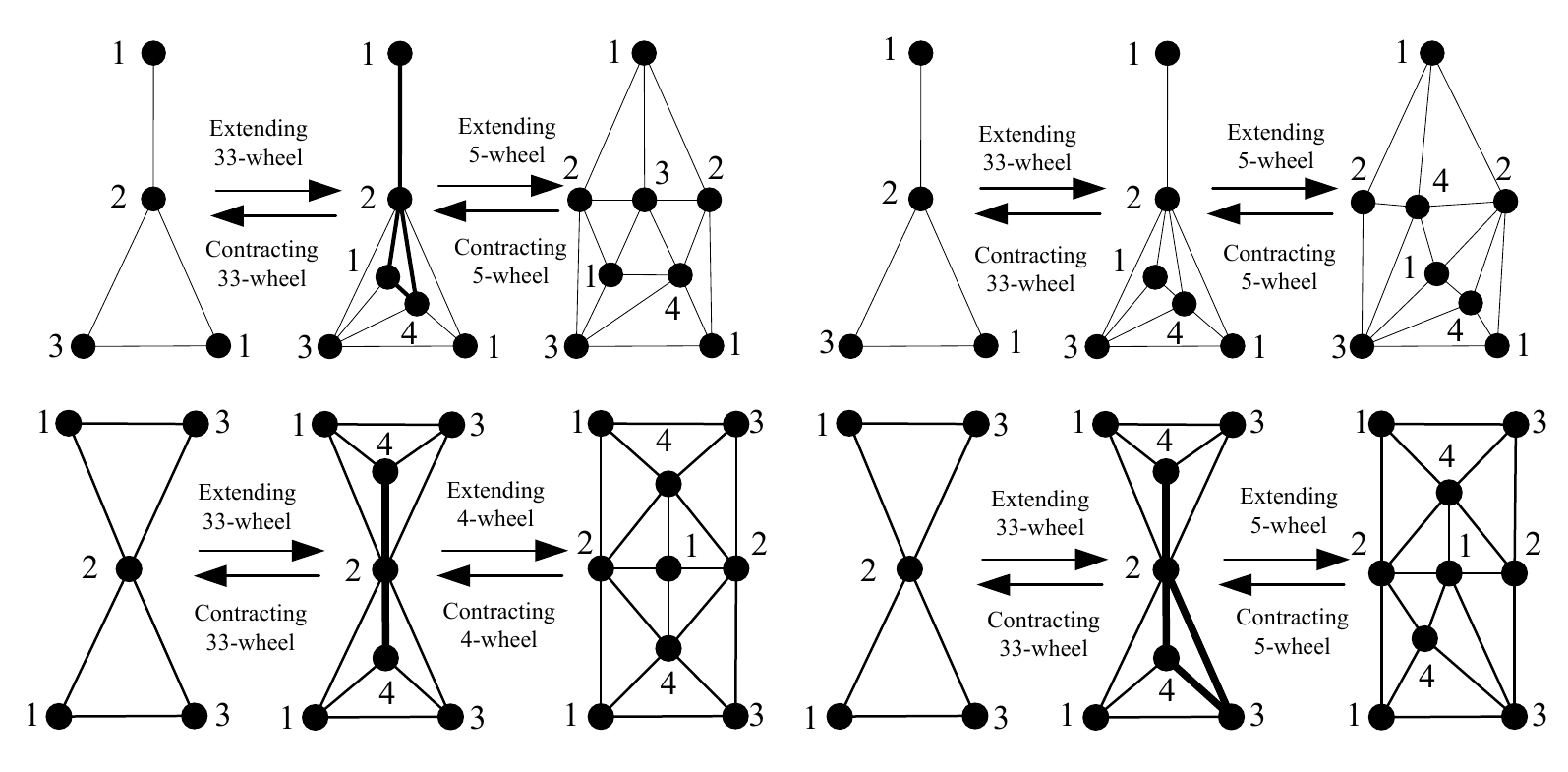}\\
\caption{\label{fig:Domino-2}{\footnotesize Another group of four domino configurations with three vertices as the centers of 3-wheel, 4-wheel and 5-wheel. }}
\end{figure}

\textbf{Method 3.} In Fig.\ref{fig:Domino-1} and Fig.\ref{fig:Domino-2}, we show Xu's Domino Extending-contracting Operational System \cite{Jin-Xu-(2)-2016}. Xu's methods can maintain the number of colors to be at most four in constructing planar graphs, and there are only two extending and contracting operations on 2-wheel, 3-wheel, 4-wheel and 5-wheel. Thereby, we say that Xu's methods are simple and easy for users to make their Topsnut-GPWs, and can make more complex Topsnut-GPWs with hundred and thousand of vertices by the Domino Extending-contracting Operational System (DE-COS). Xu's methods guarantee at most four colors are used in DE-COS.

\begin{figure}[h]
\centering
\includegraphics[height=3cm]{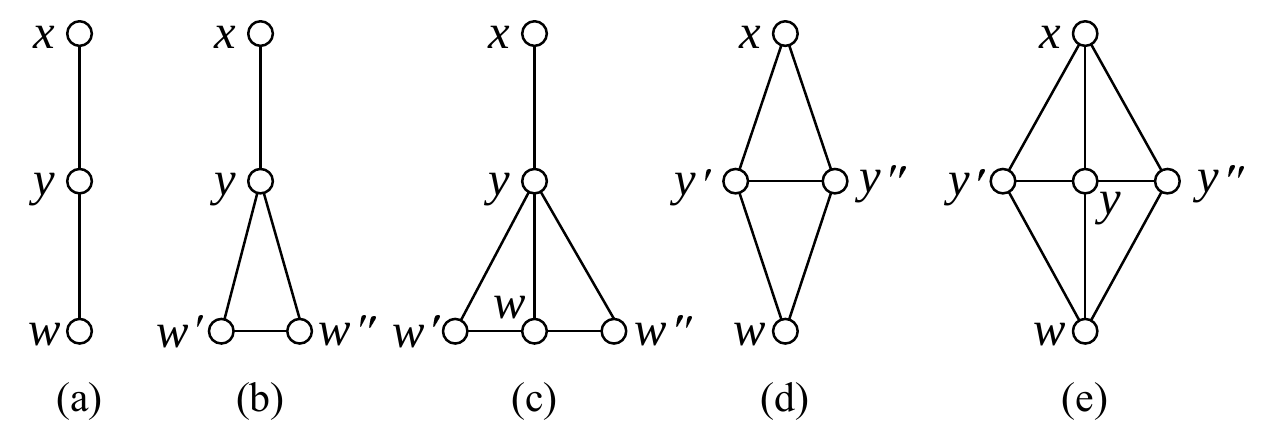}\\
\caption{\label{fig:offsprin-split-0}{\footnotesize An explanation for the offspring and splitting operations. }}
\end{figure}

\textbf{Method 4.} We define two particular operations: the \emph{offspring operation} and the \emph{splitting operation} on vertices of planar graphs as follows (ref. \cite{YAO-SUN-ZHANG-LI-YAN-2017}):

(i) In Fig.\ref{fig:offsprin-split-0} (a), we split the vertex $w$ with its neighbor set $N(w)=\{w_1,w_2,\dots ,w_d\}$ (where $w_i$ is adjacent to $w$ by the clockwise direction in the plane for keeping the planarity) into two subvertices $w',w''$ and delete the edges $ww_i$ with $1\leq i\leq d$ and $i\neq j$; next we join $w'$ with $w''$ by an edge, and join $w'$ with each of $\{w_1,w_2,\dots ,w_j\}$, and join $w''$ with each vertex of $\{w_j, w_{j+1},\dots ,w_d\}$, respectively. Here, $w_j=y$ shown in Fig.\ref{fig:offsprin-split-0} (b). The resulting graph is still a planar graph (see Fig.\ref{fig:offsprin-split-0} (b)). The above procedure is called ``doing a \emph{split operation} to the vertex $w$'' and call Fig.\ref{fig:offsprin-split-0} (b) to be a \emph{splitting graph} of Fig.\ref{fig:offsprin-split-0} (a).

Similarly, Fig.\ref{fig:offsprin-split-0} (d) is a splitting graph after doing a split operation to the vertex $y$ of Fig.\ref{fig:offsprin-split-0} (a), and we join $y'$ with each of $\{y_1,y_2,\dots ,y_j\}$; and join $y''$ with each vertex of $\{y_j, y_{j+1},\dots ,y_d,y_1\}$ with $d\geq 2$, respectively. Here, $y_j=x$ and $y_1=w$ shown in Fig.\ref{fig:offsprin-split-0} (d).

(ii) In an offspring operation, suppose that $w$ has its neighbor set $N(w)=\{w_1,w_2,\dots ,w_d\}$, where $w_i$ is adjacent to $w$ by the clockwise direction in the plane for keeping the planarity. The vertex $w$ gives birth to two vertices $w'$ and $w''$, and $w$ joins with $w',w''$ to form two edges, respectively. Next, we delete the the edges $ww_i$ with $1\leq i\leq d$ and $i\neq 1,j$, and join $w'$ with each vertex of $\{w_1,w_2,\dots ,w_j\}$, and join $w''$ with each vertex of $\{w_j, w_{j+1},\dots ,w_d\}$ respectively. Here, $w_j=y$ shown in Fig.\ref{fig:offsprin-split-0} (c). The resulting graph shown in Fig.\ref{fig:offsprin-split-0} (c), call it an \emph{offspring graph} of Fig.\ref{fig:offsprin-split-0} (a), and the process from Fig.\ref{fig:offsprin-split-0} (a) to Fig.\ref{fig:offsprin-split-0} (c) is called an offspring operation on the vertex $w$.

Furthermore, for obtaining Fig.\ref{fig:offsprin-split-0} (e), we do an offspring operation on the vertex $y$ shown in Fig.\ref{fig:offsprin-split-0} (a), where $y$ has its neighbor set $N(y)=\{y_1,y_2,\dots ,y_s\}$, where $y_i$ is adjacent to $y$ by the clockwise direction in the plane for keeping the planarity. Let $y$ join with its two birthes $y',y''$ to yield two edges, and we delete the the edges $yy_i$ with $1\leq i\leq d$ and $i\neq 1,j$ and, we join $y'$ with each vertex of $\{y_1,y_2,\dots ,y_j\}$; and join $y''$ with each vertex of $\{y_j, y_{j+1},\dots ,y_s,y_1\}$ with $d\geq 2$, respectively. Here, $y_j=x$ and $y_1=x$ shown in Fig.\ref{fig:offsprin-split-0} (e).

\begin{figure}[h]
\centering
\includegraphics[height=3cm]{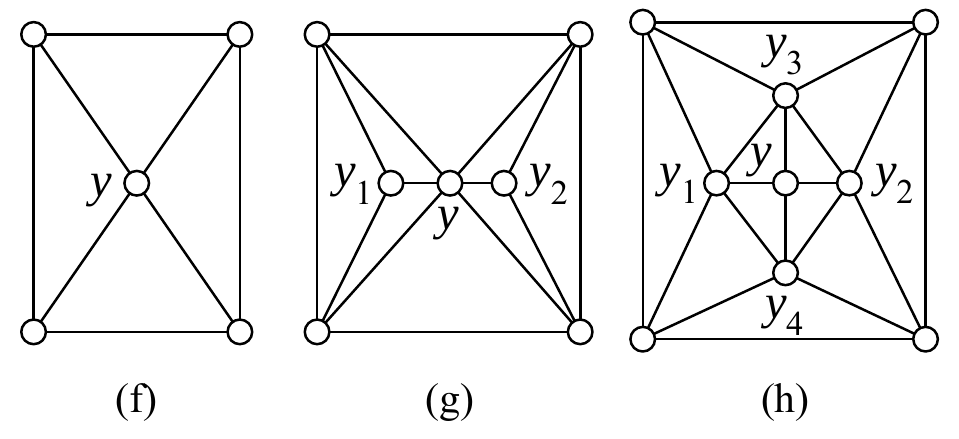}\\
\caption{\label{fig:offsprin-split-1}{\footnotesize An example obtained by doing the offspring operation twice. }}
\end{figure}

\textbf{Method 5.} The concept of the \emph{flip operation} was introduced by Wagner \cite{Wagner-1936}. In 2001, Gao \emph{et al.} \cite{Wagner-Urrutia-Wang-1936} proved that every maximal planar graph with $n$ vertices contains at least $n-2$ flippable edges; and there exist some maximal planar graphs containing at most $n-2$ flappable edges. Moreover, Gao \emph{et al.} showed that there were at least $2n + 3$ flippable edges in a maximal planar graph $G$ if $\delta(G)\geq 4$.

\begin{figure}[h]
\centering
\includegraphics[height=2.8cm]{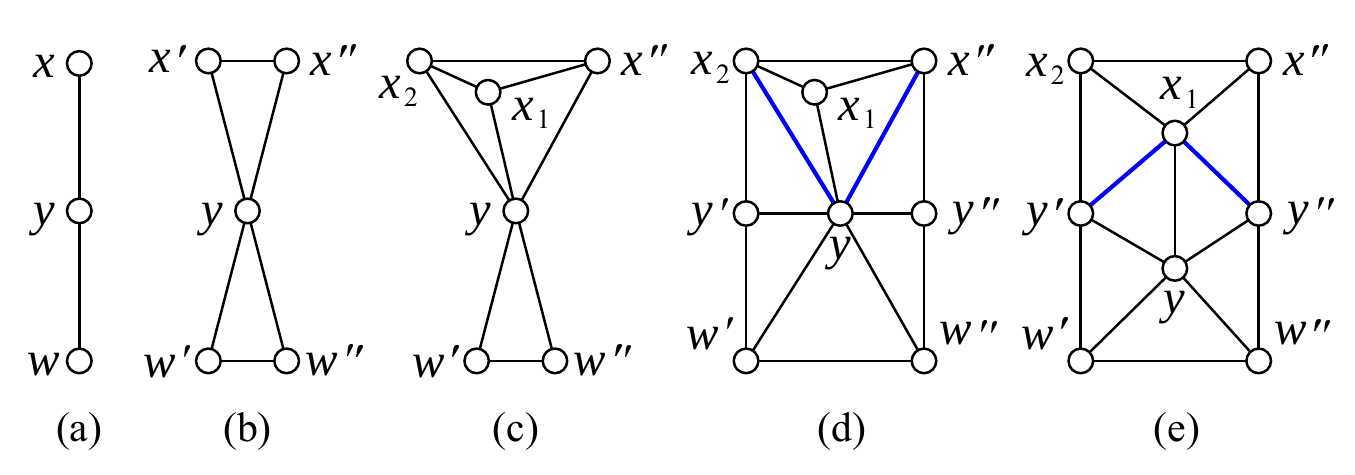}\\
\caption{\label{fig:flippable-edge-1}{\footnotesize First example for illustrating the flip operation. }}
\end{figure}

In Fig.\ref{fig:flippable-edge-1}, (b) is obtained from (a) by doing two split operations on two vertices $x,w$; and (c) is obtained from (b) by doing a split operation on vertex $x'$; (d) is obtained from (c) by doing an offspring operation on vertex $y$; and (e) is obtained from (d) by doing two flip operations on two edges $x_2y$ and $x''y$, where two flip operations are to delete the edges $x_2y$ and $x''y$, and then join vertex $y'$ with $x_1$ by an edge, next, join vertex $y''$ with $x_1$ by an edge.

\begin{figure}[h]
\centering
\includegraphics[height=2.8cm]{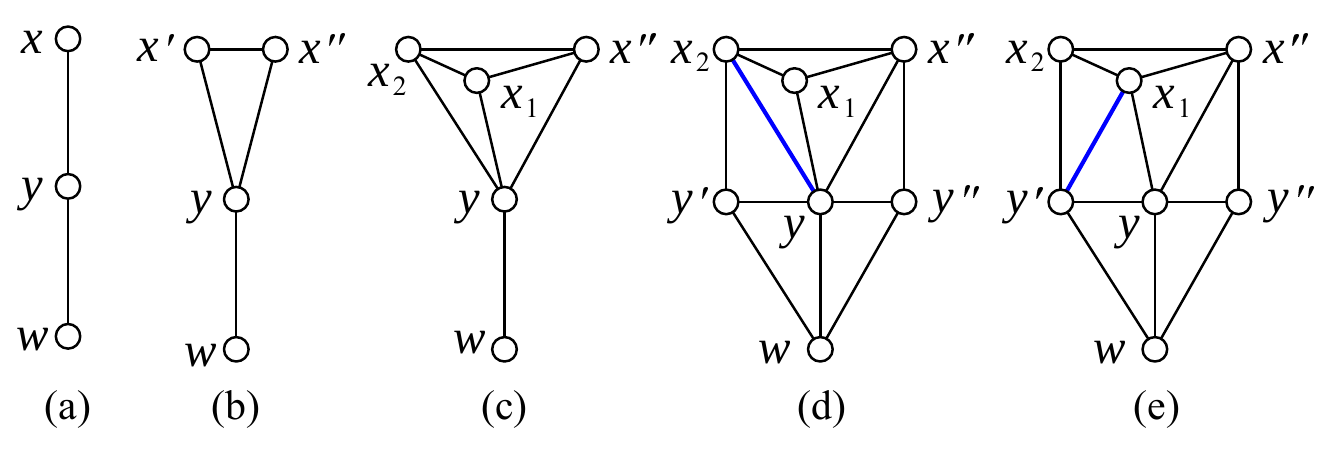}\\
\caption{\label{fig:flippable-edge-2}{\footnotesize Second example for illustrating the flip operation. }}
\end{figure}

\vskip 0.2cm

\subsubsection{Triangularly edge-identifying and edge-subdivision operations} Let $F_{\textrm{TPG}}$ be the set of planar graphs such that each one of $F_{\textrm{TPG}}$ has its outer face to be triangle and a proper 4-coloring.

\begin{figure}[h]
\centering
\includegraphics[height=2cm]{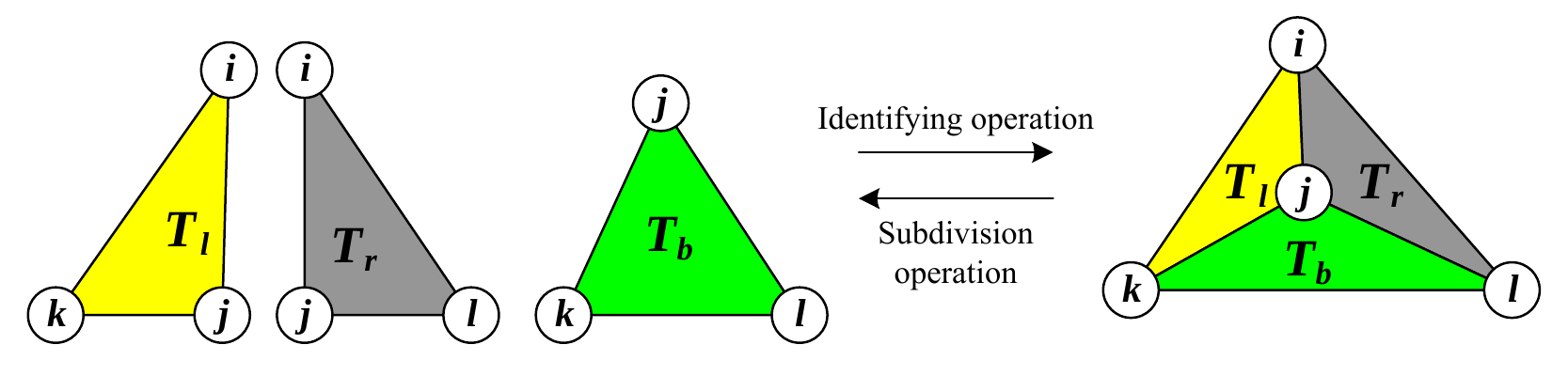}\\
\caption{\label{fig:Triangular-operation}{\footnotesize The scheme for illustrating the triangularly edge-identifying and triangular edge-subdivision operations. }}
\end{figure}

In Fig. \ref{fig:Triangular-operation}, $T_l\in F_{\textrm{TPG}}$ indicates the left planar graph having a 4-coloring $f_l$, $T_r\in F_{\textrm{TPG}}$ indicates the right planar graph having a 4-coloring $f_r$ and $T_b\in F_{\textrm{TPG}}$ is the bottom planar graph having a 4-coloring $f_b$. We identify the edge $ij$ of $T_l$ with the edge $ij$ of $T_r$ into one edge denoted as $ij$, so we get a planar graph $H$ having it outer face $kilj$, and then we identify the edge $jk$ of $T_b$ with the edge $jk$ of $H$ into one edge and identify the edge $jl$ of $T_b$ with the edge $jl$ of $H$ into one edge. Finally, we get a planar graph $G(T_l,T_r,T_b)$ having a 4-coloring obtained by three 4-colorings $f_l$, $f_r$ and $f_b$. Clearly, the planar graph $G(T_l,T_r,T_b)$ belongs to $F_{\textrm{TPG}}$. We call the above procedure of building up $G(T_l,T_r,T_b)$ a \emph{triangularly edge-identifying operation}. Conversely, we can subdivide $G(T_l,T_r,T_b)$ into $T_l,T_r$ and $T_b$, call such operation a \emph{triangular edge-subdivision operation}, also, $G(T_l,T_r,T_b)$ is \emph{triangularly edge-subdivisible}.

If the topological structures of $T_l,T_r$ and $T_b$ are isomorphic to each other, then $G(T_l,T_r,T_b)$ is \emph{triangularly regular subdivisible}.

We can use these tow operations to study some properties of maximal planar graphs.

\begin{rem} \label{rem:1111}
(1) The triangularly edge-identifying operation and triangular edge-subdivision operation will close many properties of the TEEoO-factor and the TEEoO-object, such as various 4-colorings.

(2) The triangularly edge-identifying operation and triangular edge-subdivision operation can be generalized to planar graphs having non-triangular outer faces.
\end{rem}

\subsection{Analysis of Topsnut-GPWs}

\subsubsection{A general definition for Topsnut-GPWs} We use the function concept to present a general definition of Topsnut-GPWs as follows:
\begin{defn} \label{defn:Topsnut-GPW-function}
Let $l$ stand up a lock (authentication), $k$ be a key (password), and $h$ be the password rule (a procedure of authentication). The function $l=h(k)$ represents the state of ``a key $k$ open a lock $l$ through the password rule $h$, and call directly $l=h(k)$ a \emph{Topsnut-GPW}.
\end{defn}

A Topsnut-GPW $l=h(k)$ contains three major aspects: \emph{pattern, order and structure}, also, the mathematical principles. Let $D(h)$ and $R(h)$ be the domain and the range of the function $h$, respectively. The complex of the Topsnut-GPW $l=h(k)$ is determined by the password rule $h$, the domain $D(h)$ and the range $R(h)$. If one of cardinalities of $D(h)$ and $R(h)$ is the exponential form, then the complex of the Topsnut-GPW $l=h(k)$ is not polynomial; if the password rule $h$ is NP-hard, thereby, so is the Topsnut-GPW $l=h(k)$ too.

In visualization, a Topsnut-GPW $l=h(k)$ is a labeled graph $G$ by a coloring/labelling $f$ belonging to a particular class $F$. So, $D_{dif}(G)=(p,q,r,s)$ is defined as the basic difficulty of the Topsnut-GPW $G$, where parameters $p=|V(G)|$, $q=|E(G)|$, $r$ graph properties and $s$ coloring/labellings belonging to $F\setminus \{f\}$.

\vskip 0.2cm

\subsubsection{Methods to judge different Topsnut-GPWs} 
\begin{figure}[h]
\centering
\includegraphics[height=3cm]{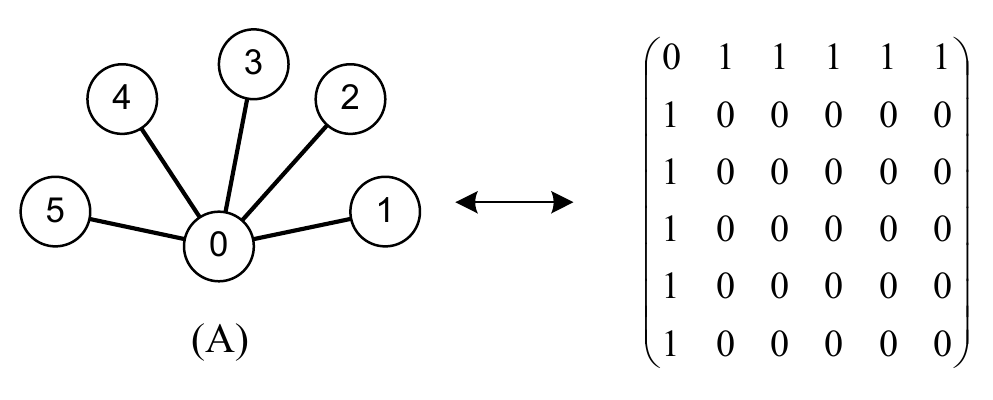}\\
\caption{\label{fig:graph-matrix-1}{\footnotesize Left (A) is shown in Fig.\ref{fig:graph-2}, right is the matrix $M(A)$ of (A). }}
\end{figure}

\begin{figure}[h]
\centering
\includegraphics[height=3cm]{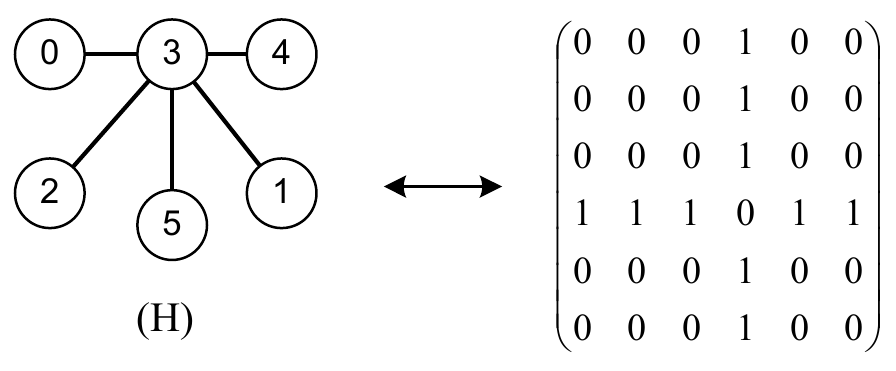}\\
\caption{\label{fig:graph-matrix-2}{\footnotesize Left (H) is shown in Fig.\ref{fig:graph-3}, right is the matrix $M(H)$ of (H). }}
\end{figure}

All topological structures used in Topsnut-GPWs are storage in computer by graph matrices. Four Topsnut-GPWs shown in Fig.\ref{fig:graph-2} correspond a graph matrix $M(A)$ shown in Fig.\ref{fig:graph-matrix-1}, and the graph matrix $M(A)$ is not equal to the graph matrix $M(H)$ shown in Fig.\ref{fig:graph-matrix-2}. Thereby, graph matrices enable us to judge different Topsnut-GPWs.

\vskip 0.2cm

\subsubsection{Non-symmetrization} Topsnut-GPWs can solve the problem of ``one key open two or more locks, and one lock can be opened by two or more keys'' (onekey-to-morelocks,~onelock-to-morekeys). In general, the onelock-to-morekeys is a function
\begin{equation}\label{eqa:c3xxxxx}
l=h(k_1, k_2, \dots, k_m),~m\geq 2,
\end{equation}
where $l$ is a lock, and each $k_i$ is a key for $i=1,2,\dots, m$. Conversely, the onekey-to-morelocks is the inverse of the onelock-to-morekeys as follows
\begin{equation}\label{eqa:c3xxxxx}
k=p^{-1}(l_1, l_2, \dots, l_n),~n\geq 2,
\end{equation}
where $k$ is a key, and each $l_j$ is a lock with $j=1,2,\dots, n$.

\begin{figure}[h]
\centering
\includegraphics[height=3.2cm]{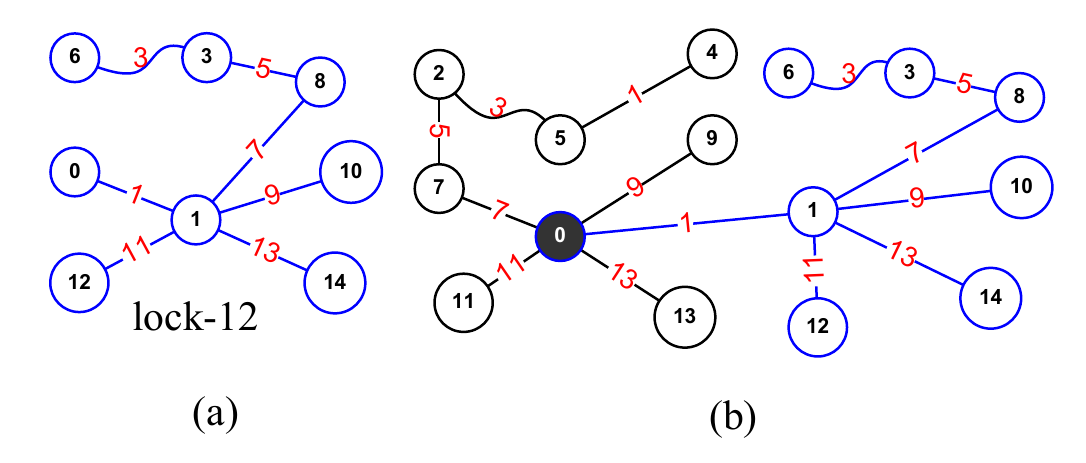}\\
\caption{\label{fig:GPW-3}{\footnotesize The key (D) shown in Fig. \ref{fig:GPW-1} has a lock (a) which differs from that shown in Fig. \ref{fig:GPW-1} (E).}}
\end{figure}

\begin{figure}[h]
\centering
\includegraphics[height=3.2cm]{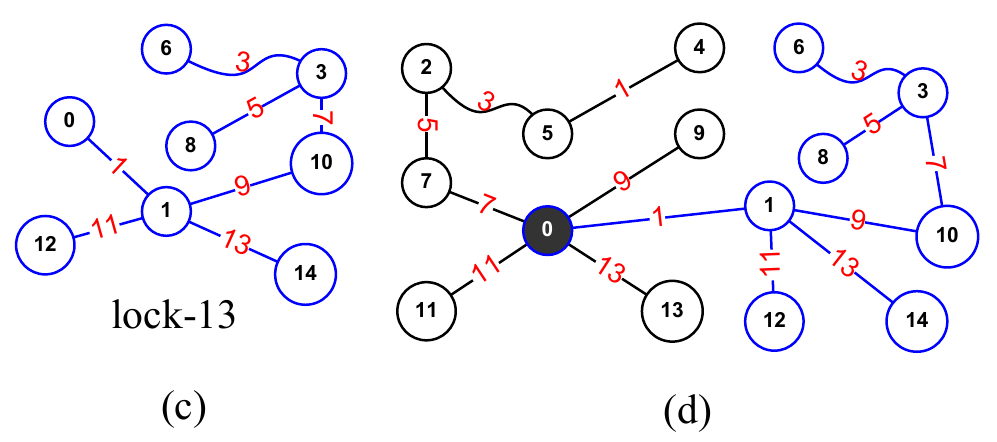}\\
\caption{\label{fig:GPW-4}{\footnotesize The key (D) shown in Fig. \ref{fig:GPW-1} has a lock (c) which differs from that shown in Fig. \ref{fig:GPW-1} (E), also differs from one shown in Fig. \ref{fig:GPW-3} (a).}}
\end{figure}

An example is shown by Fig. \ref{fig:GPW-1}, Fig. \ref{fig:GPW-3} and Fig. \ref{fig:GPW-4}. The key (D) shown in Fig. \ref{fig:GPW-1} has three locks shown in Fig. \ref{fig:GPW-1} (E), Fig. \ref{fig:GPW-3} (a) and Fig. \ref{fig:GPW-4} (c), respectively. Conversely, we can say: there are three keys (Fig. \ref{fig:GPW-1} (E), Fig. \ref{fig:GPW-3} (a) and Fig. \ref{fig:GPW-4} (c)) corresponding a lock (Fig. \ref{fig:GPW-1} (D)).

\vskip 0.2cm

\subsubsection{Topsnut-GPW chains}

We are given a sequence $K$ of Topsnut-GPWs $k_1, k_2, \dots, k_m$ such that $k_{i+1}=g(k_i)$ with $i=1, 2, \dots, m-1$, where each key $k_{i+1}$ is obtained by the key $k_{i}$, then the Topsnut-GPW $l=h(k_m)$ is called an \emph{$m$-rank Topsnut-GPW}, the sequence $K$ is called a \emph{recursive Topsnut-GPW chain}. Obviously, the greater value of $m$ and the more difficult to be break down, but the difficulty of users' memory then increases. Also, we can define another Topsnut-GPW chain by $k_{i+1}=g(k_1, k_2, \dots, k_i)$ with $i=1, 2, \dots, m-1$, or a \emph{Fibonacci Topsnut-GPW chain} defined by $k_{j+1}=g(k_{j-1}, k_j)$ with $j=2, \dots, m-1$.

\begin{rem} \label{rem:1111}
(1) It may be interesting to add the thought of \emph{Markov chain} in Topsnut-GPW chains.

(2) Recursive planar $O_r$-graphs $G_0$, $G_0,\dots ,$ $G_{n}$ form a Topsnut-GPW chains under a recursive operation $O_r$. Here, $O_r$ may be one of the triangularly edge-identifying operation, the triangularly embedded edge-overlapping operation and the triangularly single-edge-paste operation, and so on.
\end{rem}

\subsubsection{Perfect $\eta$-labeling graphs}

We consider an interesting class of graphs as studying Topsnut-GPWs, we call such particular graphs as \emph{perfect $\eta$-labeling graphs} defined by: ``Let $\eta$-labeling be a given graph labelling, and let a connected graph $G$ have a $\eta$-labeling. If every connected proper subgraph of $G$ also admits this $\eta$-labeling, then we call $G$ a \emph{perfect $\eta$-labeling graph}.'' As known, all caterpillars are (odd-)graceful, so each caterpillar is a perfect (odd-)graceful labeling graph. In fact, caterpillars admit many graph labellings. By the technique used in \cite{Zhou-Yao-Chen-Tao2012}, we can show that all lobsters are perfect (odd-)graceful labeling graphs. We ask for: \emph{If every connected proper subgraph of a connected graph $G$ has a $\eta$-labelling, then does $G$ admits this $\eta$-labelling too}? Clearly, a perfect $\eta$-labeling graph (like a mother) can be used to produce a crowd of Topsnut-GPWs (like sons and girls). So, perfect $\eta$-labeling graphs can be used to solve problems of one-key vs more-locks, or one-lock vs more-keys.

\vskip 0.2cm

\subsubsection{Connection between the Topsnut-GPWs based on a graph} Suppose that a graph $G$ admits two different labellings $f_i: X\rightarrow [a_i,b_i]$ with $i=1,2$, where $X\subseteq V(G)\cup E(G)$, and each $f_i$ holds a given restriction $c_i$ with $i=1,2$ (such as, $c_i$ is graceful, or odd-graceful, or edge-magic total and so on). So, we have two labeled graphs $G_i$ having labelling $f_i$ with $i=1,2$, and we can have a correspondence $h$ between $X\subseteq E(G)\cup V(G_1)$ and $X\subseteq V(G_2)\cup E(G)$ such that $h: f_1(x) \leftrightarrow f_2(x) $ for $x\in X$.

For example, a key $T_1$ has a labelling $f_1$ shown in Fig.\ref{fig:GPW-1}(D), and another key $T_2$ has a labelling $f_2$ shown in Fig.\ref{fig:new-example-duals}(a). So, we have a correspondence $f_1\leftrightarrow f_2$ defined by: $4\leftrightarrow 10$, $9\leftrightarrow 5$, $2\leftrightarrow 12$, $7\leftrightarrow 7$, $0\leftrightarrow 14$, $5\leftrightarrow 9$, $11\leftrightarrow 3$ and $13\leftrightarrow 1$.

\begin{rem} \label{rem:777}
A similar investigation has appeared in \cite{Yao-Liu-Yao-2017}.
\end{rem}

\subsubsection{Dual labellings of Topsnut-GPWs}

\begin{figure}[h]
\centering
\includegraphics[height=6cm]{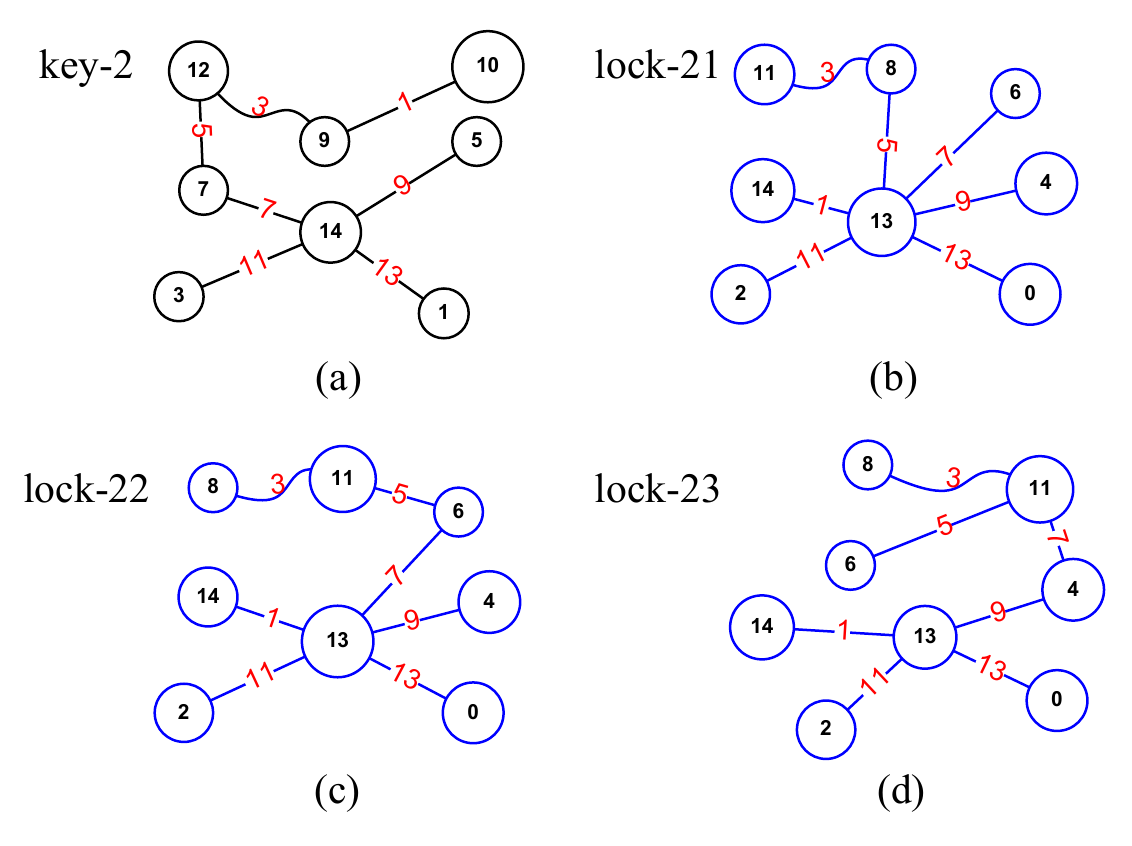}\\
\caption{\label{fig:new-example-duals}{\footnotesize (a) is the dual of the key (D) shown in Fig.\ref{fig:GPW-1}, and has three locks (b), (c) and (d).}}
\end{figure}

Many of Topsnut-GPWs made by graph labellings have their own \emph{dual labellings}. Such dual labellings can be found in literature on graph labellings. We show an example in Fig.\ref{fig:new-example-duals} here, but presenting introduction in detail.

\vskip 0.2cm

\subsubsection{Difficult rank/grades of Topsnut-GPWs}

How to determine the difficult rank/grade of a Topsnut-GPW? We have no any existing method now. Clearly, we will probe deeply Topsnut-GPWs for getting more their properties and characteristics, for example, the length of a Topsnut-GPW, the round number of authentication, non-symmetrization, Topsnut-GPW chain, etc. And we should measure these properties by mathematical techniques in order to show scientific ranks for Topsnut-GPWs.

\subsection{Topsnut-GPWs' spaces}

\vskip 0.2cm

\subsubsection{Topsnut-GPW' graph-spaces}

Wang \emph{et al.} have worked out constructions of large scale of Topsnut-GPWs by given smaller scale Topsnut-GPWs (Ref. \cite{Wang-Xu-Yao-2016, Wang-Xu-Yao-Key-models-Lock-models-2016, Wang-Xu-Yao-2017, Wang-Yao-Yang-Yang-Chen-Yao-Zhao-2013, Wang-Yao-Yang-Yang-Chen-2013, Wang-Xu-Yao-2018}). Wang \emph{et al.} have built up a Topsnut-GPW $H$-space $F$ by a given basis graph $H$ and a given group of smaller scale Topsnut-GPWs $G_1,G_2,\dots , G_n$ such that each element $G$ of $F$ is denoted as $H(G_1,G_2,\dots , G_n)$, which is just a high-level Topsnut-GPW having at least $\sum^n_{k=1}|V(G_k)|$ vertices.
We use $G_p$ to indicate the number of graphs of order $p$, then Harary and Palmer \cite{Harary-Palmer-1973} have shown the numbers of graphs of order $p$ (see Table-1)

We, often, use trees to produce Topsnut-GPWs, because trees admit many graph labellings \cite{Gallian2016}. Let $t_p$ and $T_p$ be the numbers of non-isomorphic trees and rooted trees, respectively. We have the numbers of trees of order $p\leq 24$ in the Table-2. Some particular Topsnut-GPWs need the help of digraphs (see
Table-3).

\vskip 0.2cm

\subsubsection{Coloring/labelling spaces} Sheppard \cite{Sheppard-D-A-1976} has shown that there are exactly $q!$ gracefully labeled graphs with $q$ edges, where $\frac{1}{2} q!$ of these correspond to different labellings of the same graph. No report is about the number of graph colorings or the number of graph labellings in our memory. There are over $100$ graph labellings listed in \cite{Gallian2016}.

\vskip 0.2cm

\subsubsection{Measuring Topsnut-GPWs' spaces}

A $(p,q)$-graph $G$ is a graph having $p$ vertices and $q$ edges. The saying ``distinct labellings $g$'' means the distinct labellings belong to the class that contains $g$; similarly, the sentence ``distinct colorings $f$'' means the distinct colorings belong to the class that contains $f$. However, the saying ``different type labellings (resp. colorings)'' means that a $(p,q)$-graph $G$ has all of different type labellings (resp. colorings). We define the following basic metric parameters for measuring Topsnut-GPWs' spaces:
\begin{itemize}
\item $n(p,q)$, the number of non-isomorphic $(p,q)$-graphs.

\item $\overrightarrow{n}(p,q)$, the number of non-isomorphic digraphs having $p$ vertices and $q$ arcs.

\item $n(p)$, the number of non-isomorphic graphs of order $p$.

\item $\overrightarrow{n}(p)$, the number of non-isomorphic digraphs of order $p$.

\item $n_{l}(G,g)$, the number of distinct labellings of a $(p,q)$-graph $G$ for a special labelling $g$.

\item $n_{c}(G,f)$, the number of distinct colorings $f$ of a graph $G$.

\item $n_{set}(G,h)$, the number of distinct set-colorings $h$ of a graph $G$.

\item $a_{c}(G)$, the number of different type colorings of a graph $G$. Notice that $a_{c}(G)\neq a_{c}(H)$ for two $(p,q)$-graphs $G$ and $H$, in general.

\item $a_{l}(G)$, the number of different type labellings of a graph $G$. In general $a_{l}(G)\neq a_{l}(H)$ for two $(p,q)$-graphs $G$ and $H$.

\item $a_{set}(G)$, the number of different type of set-colorings of a graph $G$. In general $a_{set}(G)\neq a_{set}(H)$ for two $(p,q)$-graphs $G$ and $H$.
\end{itemize}

We have four types of line/curve, dot-line/curve, line circles, dot-line circles in our design of Topsnut-GPWs. In a $(p,q)$-graph $G$, we have $2^p$ different line and dot-line circles, and $2^q$ different line/curves and dot-line/curves. Roughly speaking, we have $2^{p+q}$ different expressions of a $(p,q)$-graph $G$. Thereby, the $(p,q)$-graph $G$ can yield the number $M(G)$ of distinct Topsnut-GPWs is
\begin{equation}\label{eqa:space-measuring-11}
{
\begin{split}
M(G)&=2^{p+q}\cdot \big [a_{c}(G)\cdot n_{c}(G,f)+a_{l}(G)\cdot n_{l}(G,g)\\
&\quad+a_{set}(G)\cdot n_{sey}(G,h)\big ]
\end{split}}
\end{equation}
with $q=1,2,\dots, p(p-1)/2$. All $(p,q)$-graphs can produce $M(p,q)$ Topsnut-GPWs as follows
\begin{equation}\label{eqa:space-measuring-22}
{
\begin{split}
M(p,q)&=n(p,q)\cdot 2^{p+q}\cdot \big [a_{c}(G)\cdot n_{c}(G,f)\\
&\quad +a_{l}(G)\cdot n_{l}(G,g)+a_{set}(G)\cdot n_{sey}(G,h)\big ]
\end{split}}
\end{equation}
for $q=1,2,\dots, p(p-1)/2$. If we use $k_c$ colors to color lines and circles, we have $k_c^{p+q}\cdot M(G)$ Topsnut-GPWs based on a graph $G$, and $k_c^{p+q}\cdot M(p,q)$ Topsnut-GPWs based on all $(p,q)$-graphs.

We take trees of order 10 for computing $M(10,9)$. In table-2, we can see $t_{10}=106$ and $T_{10}=719$. So, by the result due to Sheppard \cite{Sheppard-D-A-1976}, each tree of order 10 has $\frac{1}{2} 10!$ graceful labellings, that is, $n_{l}(G,g)=10!/2$. We have
$${
\begin{split}
M(10,9)&=t_{10}\cdot2^{p+q}\cdot n_{l}(G,g)=106\cdot 2^{18}\cdot  10!\\
&=100834423603200\approx 2^{46.51898157}
\end{split}}
$$
For all rooted trees of order 10, we have
$${
\begin{split}
M_{rooted}(10,9)&=T_{10}\cdot2^{p+q}\cdot n_{l}(G,g)=719\cdot 2^{18}\cdot 10!\\
&=683961797836800.00 \approx 2^{49.28090908}
\end{split}}
$$

As comparing, roughly saying, the earth sand amount is about $8\cdot 10^{22}$, or $13\cdot 10^{23}$ ($2^{76.08241809}\sim 2^{76.78285781}$), which is 8 trillion to 13 trillion billion billion; and the number of stars in the most sophisticated telescope can be observed is about $7\cdot 10^{22}$ ($2^{75.88977301}$). Suppose that there are ten billion people in the world, then so each person has $M(10,9)\div$ten billion$\approx 100834$ Topsnut-GPWs; or for one hundred billion people, thus, each person has $M(10,9)\div$one hundred billion$\approx 1008$ Topsnut-GPWs.

Since $G_{24}$ in Table-1 consist of 60-figure numbers, we can label a graph of order 24 by two or more styles, that means we can get digital passwords having more than 120-figures. In other words, we will wait for Xu's computer \cite{Jin-Xu-Probe-Machine-2016}, \emph{Probe Machine computer}, to break down our Topsnut-GPWs.

\section{Conclusion and further researches}

We have shown the constitution of the existing GPWs, and list several possible attack methods on the existing GPWs. We enjoy the Topsnut-GPWs made by the idea proposed by Wang \emph{et al.}, and focus on the Topsnut-GPWs used by mobile devices with touch screen. We can see that Topsnut-GPWs having personal customization, high cultural degree and intelligence will also appear and gradually replace, update, and improve the existing passwords. And the password flows and password groups will run on information networks. We say our Topsnut as a platform, called \emph{Topsnut-platform}, and we consider Topsnut-GPWs as \emph{mathematical fingerprints}. We can put some things into circles or lines on \emph{Topsnut-platform}, and then join some pairs of circles to form a story having scientific techniques or life knowledge, and so on.

\subsection{Commercial GPWs} In \cite{Gao-Jia-Ye-Ma-2013}, the authors mentioned: To date, there are only two commercial products of Drawmetric graphical password scheme. (1) An unlock scheme resembling a mini Pass-Go has been used to unlock screens on Google Android cell phones. (2) In the Window 8 system, Microsoft introduces a new graphical password. Where, V-GO is a commercial graphical password scheme developed by Passlogix based on Blonder's idea. A similar technique, visKey, was developed by Sfr, and is a commercial version of PassPoints for the PPC (Pocket Personal Computer). This scheme is used for screen-unlock by tapping on a correct sequence of click-points with a stylus or finger. VisKey PPC combines easy handling with high security for mobile
devices. Just a few clicks in a picture may offer a large theoretical password space. GrIDsure, a commercial product, is a graphical one-time PIN scheme, which makes PINs more resistant to shoulder-surfing attacks by using graphical passwords on a grid.

Gao \emph{et al.} summarized: The two products of the Drawmetric graphical password scheme demonstrate clearly that commercial product schemes must be easy to remember, simple to operate, and apply to systems which require low security level.

\subsection{More elements for innovating Topsnut-GPWs}

For the reason of revealing individuality, we can consider the following elements when producing Topsnut-GPWs.

\begin{figure}[h]
\centering
\includegraphics[height=3.2cm]{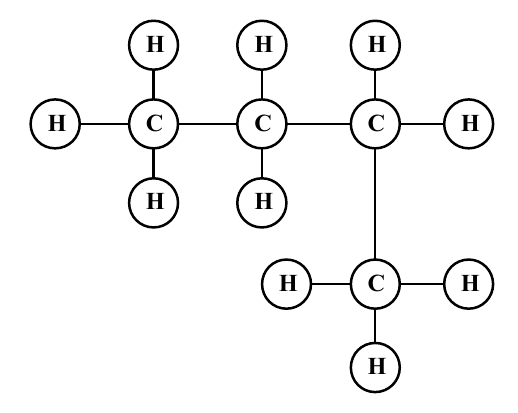}~\includegraphics[height=2cm]{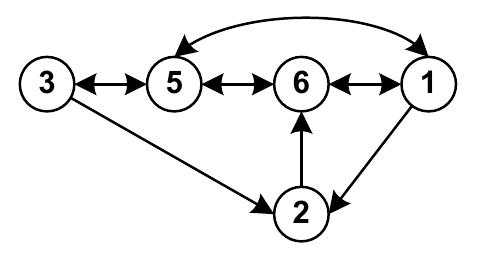}\\
\caption{\label{fig:liangzhu}{\footnotesize Left is a saturated alkane; Right is a Topsnut-GPW obtained from a segment music is a labeled digraph.}}
\end{figure}

\begin{asparaenum}[B-1.]
\item Use rectangle, triangle, star, polygon to enrich ends of vertices.

\item Make topological structures by words (\cite{Yao-Sun-Zhang-Mu-Wang-Xu-2017}), see two \emph{Chinese character graphs} (Hanzi-graphs) shown in Fig.\ref{fig:chinese-graph}, and label these topological structures with numbers to devote interesting Topsnut-GPWs (see two examples shown in Fig. \ref{fig:chinese-graph-1}). We can prove that each tree can be decompose into Hanzi-graphs. However, we use a symbol $n_{cc}(T)$ to denote the smallest number of Hanzi-graphs that can be assembled into any tree $T$ having at least one edge, and want to compute the exact value of $n_{cc}(T)$.

\begin{figure}[h]
\centering
\includegraphics[height=2cm]{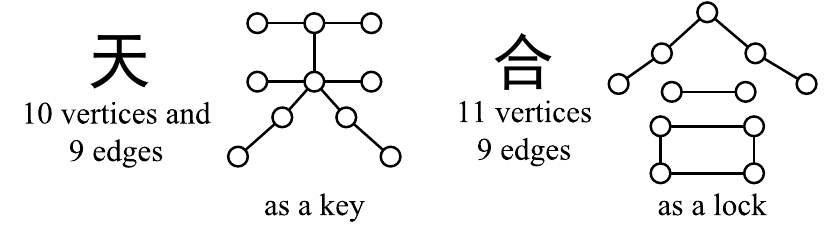}\\
\caption{\label{fig:chinese-graph}{\footnotesize Tow Hanzi-graphs made by two Chinese characters.}}
\end{figure}

\begin{figure}[h]
\centering
\includegraphics[height=2.2cm]{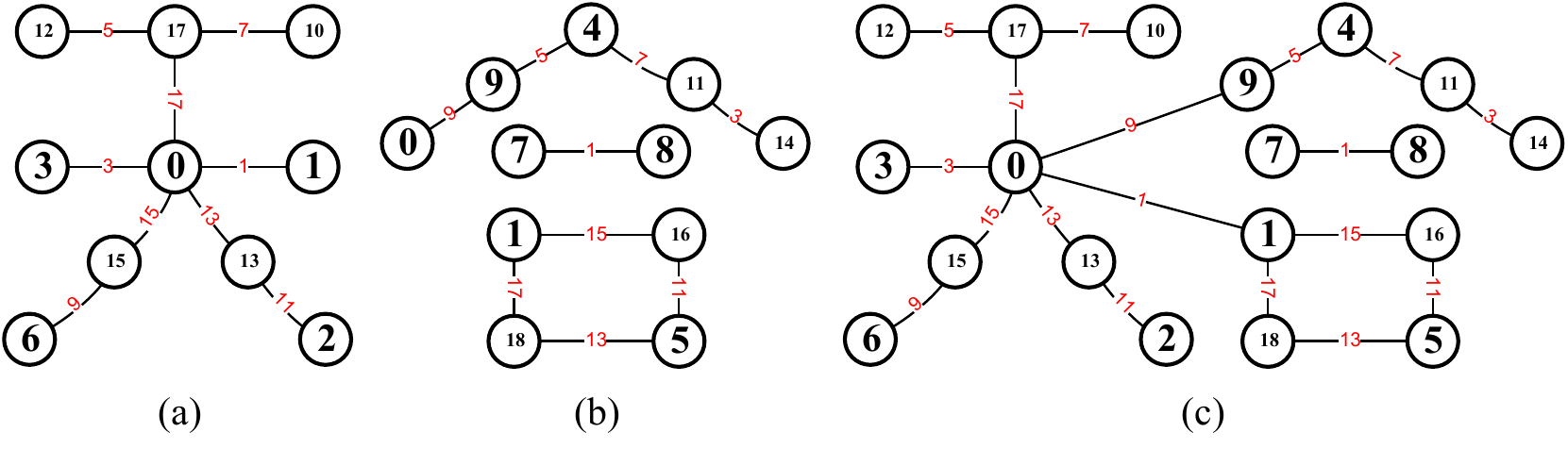}\\
\caption{\label{fig:chinese-graph-1}{\footnotesize Tow Topsnut-GPW authentications based on two Hanzi-graphs (key and lock) shown in Fig.\ref{fig:chinese-graph}. (a) the key with an odd-graceful labelling; (b) the lock  with an odd-graceful labelling; (c) an authentication.}}
\end{figure}

\item Use more mathematical knowledge in designing Topsnut-GPWs. For example, we can label small circles and lines/curves with functions, such as $\sin x$, $\cos x$, $ax^2+bx+c$, etc.
\item Topological structures can be used to describe chemical structures, in other words, Topsnut-GPWs may be related with materials (see the left picture in Fig.\ref{fig:liangzhu}).
\item Use non-mathematical methods to designing Topsnut-GPWs. For example, numbered musical notation can form a Topsnut-GPW. In the song of Butterfly Lovers (also, Liangzhu, a beautiful story of Chinese love), the first three sections contain numbers $35612615516532$, so we have a Topsnut-GPW shown in the right picture in Fig.\ref{fig:liangzhu}, in which there is a directed chain $3\rightarrow 5\rightarrow 6\rightarrow 1\rightarrow 2\rightarrow 6\rightarrow 1\rightarrow 5^2\rightarrow 1\rightarrow 6\rightarrow 5\rightarrow 3\rightarrow 2$.
\item Label small circles and lines/curves with letters such that the Topsnut-GPWs forms a poem, an interesting sentence, and so on.
\end{asparaenum}

\subsection{Complex Topsnut-GPWs}
\begin{figure}[h]
\centering
\includegraphics[height=3cm]{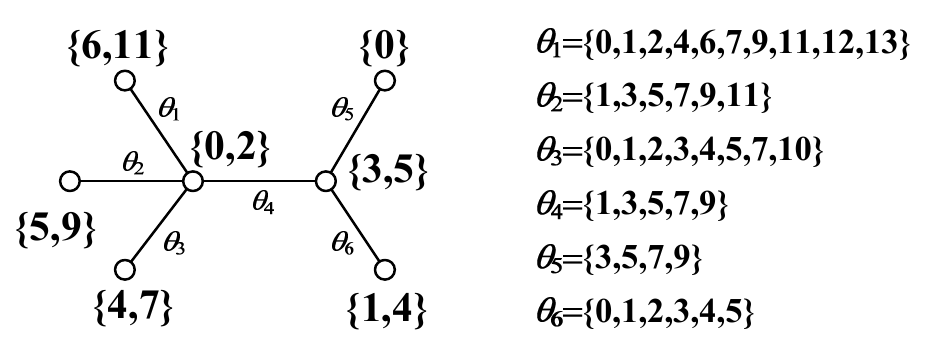}\\
\caption{\label{fig:set-coloring-labelling}{\footnotesize A complex Topsnut-GPW was obtained by graph set-colorings or set-labellings.}}
\end{figure}

In \cite{Yao-Sun-Zhang-Li-Zhao-2017} the authors introduce graph set-colorings and graph set-labellings for designing Topsnut-GPWs. A set-labelling of a tree is shown in Fig.\ref{fig:set-coloring-labelling}.

\begin{figure}[h]
\centering
\includegraphics[height=2.6cm]{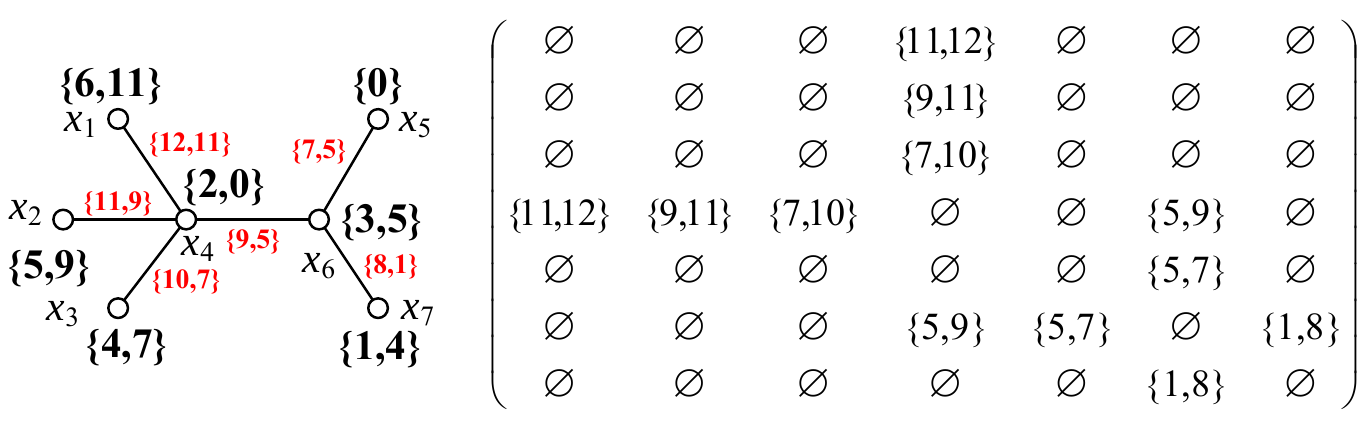}\\
\caption{\label{fig:set-matrix}{\footnotesize The left is a Topsnut-GPW $T$ obtained by set-colorings, and the right is a set-matrix obtained by the set-coloring on the edges of $T$.}}
\end{figure}

The problem of graph set-colorings/labellings yields the so-called set-matrices in
these set-matrices each element is a set only (see Fig.\ref{fig:set-matrix}).

\subsection{Biometric authentication} The so-called "biometric" refers to the combination of computer and high-tech, the use of human inherent physiological characteristics (such as fingerprint, face and iris image, etc.) and behavioral characteristics (such as handwriting, voice, gait) for personal identification. At present, the fingerprint identification is more common.

\subsection{Related with mathematical problems}

Is there a password that cannot be deciphered (broken down)? In other words, are there mathematical conjectures or mathematical problems that can not be proved?
\begin{asparaenum}[E-1.]
\item Applied software. We believe that one can design some GPW application software that has very smaller volume, and can be planted into encrypted electronic documents in order to decrypt the encrypted electronic documents by users' GPWs such that there is not needing to instal such softwares in mobile phones, iPads, and computers for reading users' GPWs.

\item Unknown parameters. In the equations (\ref{eqa:space-measuring-11}) and (\ref{eqa:space-measuring-22}), many parameters are not determined since we do not know the numbers of colorings and labellings now, and it is not easy to compute $a_{c}(G)$ and $n_{c}(G,f)$ for a given $(p,q)$-graph $G$ having a particular coloring $f$.
\item Unknown labellings determined to graphs. Most labellings are not determined to graphs, even trees and simpler graphs.

\item Unknown connections between labellings of a graph. It is not easy to find some connections between labellings of a graph, except special trees \cite{Yao-Liu-Yao-2017}. In fact, we do not know how many graph labellings dose a given graph have, and furthermore we do not know all graph labellings at all.

\item Onekey-to-morelocks,~onelock-to-morekeys. For a given key Topsnut-GPW, determine all of lock Topsnut-GPWs opened by this key; conversely, find all possible key Topsnut-GPWs to open a given lock Topsnut-GPW.

\item Set-matrices. Define operations to set-matrices, and find applications for them (see Fig.\ref{fig:set-matrix}).

\item Conjectures: We list several long-time conjectures of graph colorings/labellings in the following:

\begin{conj}\label{conj:Behzad}
1. (Behzad, 1965) Total coloring conjecture: $\chi''(G)\leq \Delta(G)+2$.

2. \emph{(Alexander Rosa, 1966) \cite{Rosa}} Each tree is graceful.

3. \emph{(Bermond, 1979) \cite{Bermond}} Every lobster is graceful.

4. \emph{(Truszczy\'{n}ski, 1984)\cite{M-Truszczynski}} All connected
unicyclic graphs, except $C_n$ for $n = 1$ or $2$ $(\emph{mod}\ 4)$,
are graceful.

5. \emph{(R.B. Gnanajothi, 1991) \cite{R-B-Gnanajothi}} Every tree is
odd-graceful.

6. (Burris and Schelp, 1993, 1997) Vertex distinguishing edge coloring conjecture.

7. (Zhang Zhongfu, Liu Linzhong Wang Jianfang, 2002) Adjacent vertex distinguishing edge coloring conjecture.

8. (Zhang \emph{et al.}, 2008) Adjacent vertex distinguishing total coloring conjecture.

9. Unique 4-color maximal plane graph conjecture, 9-color conjecture \cite{Jin-Xu-(1)-2016}.

10. \emph{(Bing Yao, 2005) \cite{Gallian2016}} The odd-gracefulness of trees is equivalent to
the gracefulness of trees.

11. (Yang \emph{et al.}, 2016) \cite{Yang-Yao-Ren-2016} A graph $G$ having a proper total colorings with four distinguishing constraints holds $\chi''_{4as}\leq \Delta(G)+4$.
\end{conj}

\quad The conjecture ``\emph{Each tree is graceful}'' is a famous graceful tree conjecture (GTC). However, GTC is open now, only few classes of graphs were verified to support GTC. It seems to be very difficult to show a graph having no graceful labelling.
Zhou et al. \cite{Zhou-Yao-Chen-Tao2012} have proven: \emph{Every lobster is odd-graceful}.

\begin{conj}\label{conj:velocity}
A maximal planar graph $G$ is 4-colorable if and only if four labelled triangles shown in Figure \ref{fig:a-conjecture} can tile fully $G$.
\end{conj}

\begin{figure}[h]
\centering
\includegraphics[height=3.2cm]{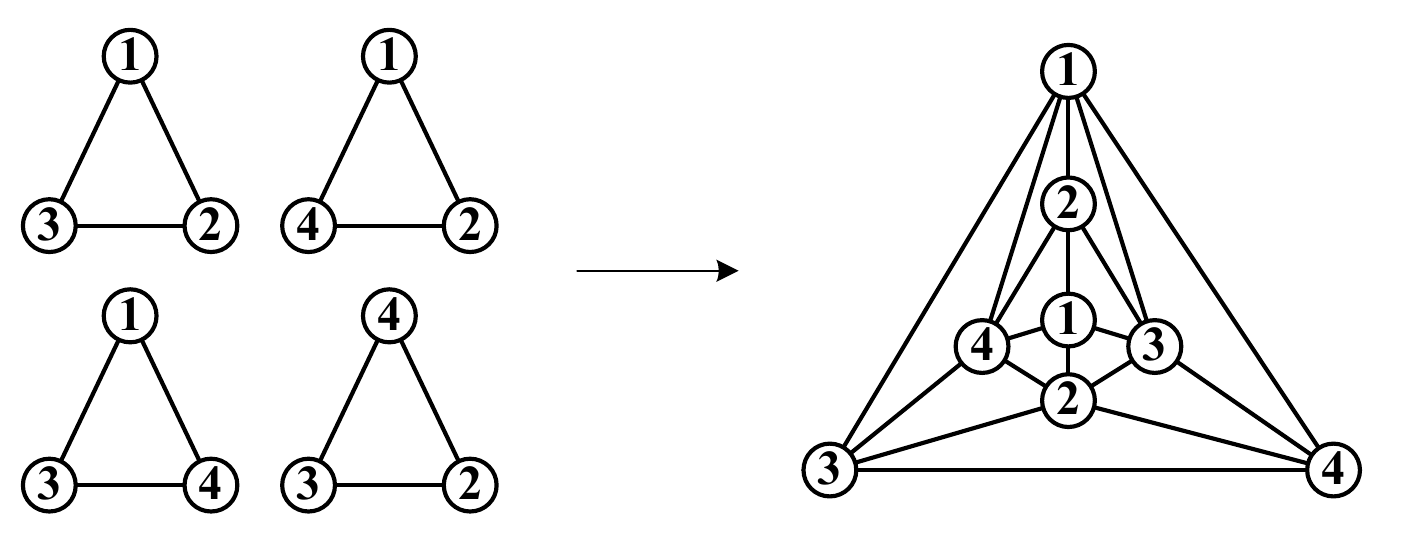}\\
\caption{\label{fig:a-conjecture}{\footnotesize  A maximal planar graph (right) tiled by four labelled triangles.}}
\end{figure}
\end{asparaenum}

\subsection{The power law in passwords}

Let $x$ be the number of the passwords, and let $y$ be the number of people who can remember their passwords. We guess that $x$ and $y$ obey the scale-free distribution, that is, we have the power law
\begin{equation}\label{eqa:power-law00}
y=cx^{-\lambda},~\lambda>0.
\end{equation}
Furthermore, we can get $yx^{\lambda}=c$ with $\lambda>0$, which is a balanced equation. However, no proof is reported for the equation (\ref{eqa:power-law00}) in our memory.

It has been mentioned in \cite{Wiedenbeck-Waters-Birget-Brodskiy-Memon-2005} that the power law of forgetting describes rapid forgetting soon after learning, followed by very slow drop-off thereafter (Bahrick, 1984; Wixted and Ebbesen, 1991). And most GPW schemes fall along the descending line in Eq. (\ref{eqa:power-law00}), where increased security $x$ implies decreased usability $y$. System-assigned passwords are generated randomly to preclude attacks exploiting skewed distributions and use larger portions of the theoretical password space, but have high usability costs: longer training times or increased likelihood that users forget passwords.

We, finally, claim that \emph{every thing can become a password}, and \emph{it is just beginning for Topsnut-GPWs}, except  quantum networks  in future.


\section{Acknowledgment}

We are grateful to the anonymous referees for their valuable and helpful comments which lead to the improvement of the paper. The author, \emph{Bing Yao}, gratefully thank the National Natural Science Foundation of China under grants No. 61163054, No. 61363060 and No. 61662066 for supporting our research on new type of graphical passwords. The author, \emph{Jin Xu}, is supported by National Key R\&D Program of China (No. 2016YFB0800700), National 973 Program of China (2013CB329600)




\begin{thebibliography}{1}

\bibitem{Suo-Zhu-Owen-2005} Xiaoyuan Suo, Ying Zhu, G. Scott. Owen. Graphical Password: A Survey. In: Proceedings of Annual
Computer Security Applications Conference (ACSAC), Tucson, Arizona. IEEE (2005) 463-472. (10 pages, 38 reference papers)

\bibitem{Biddle-Chiasson-van-Oorschot-2009}R. Biddle, S. Chiasson, and P.C. van Oorschot. Graphical passwords: Learning from the First Twelve Years. ACM Computing Surveys, 44(4), Article 19:1-41. Technical Report TR-09-09, School of Computer Science, Carleton University, Ottawa, Canada. 2009. (25 pages, 145 reference papers)

\bibitem{Gao-Jia-Ye-Ma-2013}Haichang Gao, Wei Jia, Fei Ye and Licheng Ma. A Survey on the Use of Graphical Passwords in Security. Journal Of Software, Vol. 8, No. 7, July 2013, 1678-1698. (21 pages, 88 reference papers)

\bibitem{Gallian2016} Joseph A. Gallian. A Dynamic Survey of Graph Labeling. \emph{The electronic journal of
combinatorics}, 17 (2016), \# DS6. (440 pages, 2265 reference papers)

\bibitem{Wiedenbeck-Waters-Birget-Brodskiy-Memon-2005}Susan Wiedenbeck, Jim Waters, Jean-Camille Birget, Alex Brodskiy, Nasir Memon. PassPoints: Design and longitudinal evaluation of a graphical password system. Int. J. Human-Computer Studies 63 (2005) 102-127.
\bibitem{Huanyu-Zhao-Xiaolin-Li-2007} Huanyu Zhao and Xiaolin Li. S3PAS: A Scalable Shoulder-Surfing Resistant Textual-Graphical Password Authentication Scheme. in 21st International Conference on Advanced Information Networking and Applications Workshops (AINAW'07). IEEE 2007, vol. 2. Canada, 2007, pp. 467-472.
\bibitem{Blonder-No-one-1996}G.E. Blonder. Graphical passwords. United States Patent 5559961, 1996.



\bibitem{Bondy-2008} J. A. Bondy, U. S. R. Murty. Graph Theory. Springer London, 2008.

\bibitem{Wang-Xu-Yao-2016} Hongyu Wang, Jin Xu, Bing Yao. Exploring New Cryptographical Construction Of Complex Network Data. IEEE First International Conference on Data Science in Cyberspace. IEEE Computer Society, (2016):155-160.

\bibitem{Wang-Xu-Yao-Key-models-Lock-models-2016}Hongyu Wang, Jin Xu, Bing Yao. The Key-models And Their Lock-models For Designing New Labellings Of Networks.Proceedings of 2016 IEEE Advanced Information Management, Communicates, Electronic and Automation Control Conference (IMCEC 2016) 565-5568.

\bibitem{Wang-Xu-Yao-2017}Hongyu Wang, Jin Xu, Bing Yao. Twin Odd-Graceful Trees Towards Information Security. Procedia Computer Science 107 (2017) 15-20, DOI: 10.1016/j.procs.2017.03.050

\bibitem{Jansen-Gavrila-Korolev-Ayers-Swanstrom-2003}W. Jansen, S. Gavrila, V. Korolev, R. Ayers, and R. Swanstrom. Picture Password: A Visual Login Technique for Mobile Devices, NIST Report-NISTIR7030, 2003.

\bibitem{Takada-Koike-2003}T. Takada, and H. Koike. Awase-E: Image-based authentication for mobile phones using user's favorite images. Human-Computer Interaction with Mobile Devices and Services, vol. 2795, pp. 347-351, Springer-Verlag, GmbH, 2003.


\bibitem{Dunphy-Heiner-Asokan-2010}P. Dunphy, A. P. Heiner, and N. Asokan. A closer look at recognition-based graphical passwords on mobile devices. In ACM Symposium on Usable Privacy and Security (SOUPS), July 2010.

\bibitem{Xiaoyuan-Suo-2014}Xiaoyuan Suo. A Study of Graphical Password for Mobile Devices. G. Memmi and U. Blanke (Eds.): MobiCASE 2013, LNICST 130, pp. 202-214, 2014. Institute for Computer Sciences, Social Informatics and Telecommunications Engineering 2014.

\bibitem{Jin-Xu-Probe-Machine-2016}Jin Xu, Probe Machine. IEEE Transactions on Neural Networks and Learning Systems, 2016, 27(7): 1405-1416.
\bibitem{Sheppard-D-A-1976}Sheppard, D.A. The factorial representation of balanced labeled graphs. Discrete Math. 15 (1976), 379-388.

\bibitem{Harary-Palmer-1973}Harary F. and Palmer E. M. Graphical enumeration. Academic Press, 1973.


\bibitem{Jin-Xu-(1)-2016}Jin Xu. Theory on Structure and Coloring of Maximal Planar Graphs: (1) Recursion Formulae of Chromatic Polynomial and Four-Color Conjecture. Journal of Electronics and Information Technology. Vol.38 No.4, Jul. 2016, 763-770.
\bibitem{Jin-Xu-(2)-2016}Jin Xu. Theory on Structure and Coloring of Maximal Planar Graphs: (2) Domino Configurations and Extending-Contracting Operations. Journal of Electronics and Information Technology. Vol.38 No.6, Jul. 2016, 1271-1327.
\bibitem{Jin-Xu-(3)-2016}Jin Xu. Theory on Structure and Coloring of Maximal Planar Graphs: (3) Purely Tree-colorable and Uniquely 4-colorable Maximal Planar Graph Conjectures. Journal of Electronics and Information Technology. Vol.38 No.6, Jul. 2016, 1329-1353.
\bibitem{Jin-Xu-(4)-2016}Jin Xu. Theory on Structure and Coloring of Maximal Planar Graphs: (4) $\sigma$-Operations and Kempe Equivalent Classes. Journal of Electronics and Information Technology. Vol.38 No.7, Jul. 2016, 1558-1585.


\bibitem{Rosa} Alexander Rosa. On certain valuations of the vertices of
a graph. \emph{Theory of Graphs} (International Symposium in Rome in
July 1966) pp. 349-355. Eds. \emph{Gordon and Breach}, New York;
\emph{Dunod}, Paris, 1967.

\bibitem{Bermond} J. C. Bermond. Graceful Graphs. Radio Antennae and French
Windmills. \emph{Graph Theory and Combinatorics}, pp. 13-37, Pitman,
London, 1979.

\bibitem{M-Truszczynski}M. Truszczy\'{n}ski. Graceful unicyclic graphs. \emph{Demonstatio
Mathematica}, 17 (1984), 377-387.

\bibitem{R-B-Gnanajothi} R.B. Gnanajothi. Topics in Graph Theory. Ph. D. Thesis, \emph{Madurai
Kamaraj University}, 1991.

\bibitem{Yang-Yao-Ren-2016}Chao Yang, Bing Yao, Han Ren. A Note on Graph Proper Total Colorings with Many Distinguishing Constraints. Information processing letters V 16, 6 396-400. (2016)

\bibitem{Wang-Yao-Yang-Yang-Chen-Yao-Zhao-2013}Hongyu Wang, Bing Yao, Chao Yang, Sihua Yang, Xiang'en Chen, Ming Yao, Zhenxue Zhao. Edge-Magic Total Labellings Of Some Network Models. Applied Mechanics and Materials Vols. 347-350 (2013) 2752-2757. DOI:10.4028/www.scientific.net/AMM.347-350.2752.
\bibitem{Wang-Yao-Yang-Yang-Chen-2013}Hongyu Wang, Bing Yao, Chao Yang, Sihua Yang, Xiang'en Chen. Labelling Properties Of Models Related with Complex Networks Based On Constructible Structures. Advanced Materials Research Vols. 765-767 (2013) 1118-1123. DOI:10.4028/www.scientific.net/AMR.765-767.1118.
\bibitem{Wang-Xu-Yao-2018}Hongyu Wang, Jin Xu, Bing Yao. On Generalized Total Graceful labellings of Graphs. Ars Combinatoria, volume 139 July (2018). accepted
\bibitem{Wagner-1936}Wagner K. Bemerkungen zum vierfarbenproblem. Jahresbericht der Deutschen Mathematiker-Vereinigung, 1936, 46: 26-32.
\bibitem{Wagner-Urrutia-Wang-1936} Gao Z C, Urrutia J, and Wang J Y. Diagonal flips in labeled planar triangulations. Graphs and Combinatorics, 2004, 17(4): 647-656. doi: 10.1007/s003730170006.)

\bibitem{bing-yao-others-2017-tianjin-university}Bing Yao, Hui Sun, Xiaohui Zhang, Jingwen Li, Mingjun Zhang, Jianmin Xie, Ming Yao. Applying graph theory to graphical passwords. 2017 Academic Annual Conference, Specialized Committee Of Graph Theory And System Optimization, Chinese Society Of Electronics, Circuits And Systems, Tianjin University, August 12-13, 2017.
\bibitem{Yao-Liu-Yao-2017}Bing Yao, Xia Liu and Ming Yao. Connections between labellings of trees. Bulletin of the Iranian Mathematical Society, ISSN: 1017-060X (Print) ISSN: 1735-8515 (Online), Vol. 43 (2017), 2, pp. 275-283.
\bibitem{Yao-Sun-Zhang-Li-Zhao-2017}Bing Yao, Hui Sun, Xiaohui Zhang, Jingwen Li, Meimei Zhao. Applying Graph Set-Labellings Having Constraint Sets Towards New Graphical Passwords. (2017) submitted
\bibitem{YAO-SUN-ZHANG-LI-YAN-2017}Bing Yao, Hui Sun, Xiaohui Zhang, Jingwen Li, Guanghui Yan. Graph Theory Towards Designing Graphical Passwords For Mobile Devices. submitted to 2017 IEEE 2nd Information Technology, Networking, Electronic and Automation Control Conference, (2017)
\bibitem{Zhou-Yao-Chen-Tao2012}Xiangqian Zhou, Bing Yao, Xiang-en Chen and Haixia Tao. A proof to the
odd-gracefulness of all lobsters. Ars Combinatoria \textbf{103} (2012), 13-18.
\bibitem{Yao-Sun-Zhao-Li-Yan-2017}Bing Yao, Hui Sun, Meimei Zhao, Jingwen Li, Guanghui Yan. On Coloring/Labelling Graphical Groups For Creating New Graphical Passwords. submitted (2017)

\bibitem{Sun-Zhang-ZHAO-Yao-2017}Hui Sun, Xiaohui Zhang, Meimei ZHAO, Bing Yao. New Algebraic Groups Produced By Graphical Passwords Based On Colorings And Labellings. submitted (2017)
\bibitem{Yao-Sun-Zhang-Mu-Wang-Xu-2017}Bing Yao, Hui Sun, Xiaohui Zhang, Yarong Mu, Hongyu Wang, Jin Xu.  New-type Graphical Passwords Made By Chinese Characters With Their Topological Structures. submitted (2017)
\end{thebibliography}
%


\textbf{Appendix}

\begin{flushleft}
{\footnotesize
\textbf{Table-1.} The number $G_p$ of graphs of order $p$ \cite{Harary-Palmer-1973}.
\begin{tabular}{cll}
$p$&$G_p$&bits\\
6&156&7\\
7&1044&10\\
8&12346&14\\
9&274668&18\\
10&12005168&24\\
11&1018997864&30\\
12&165091172592&37\\
13&50502031367952&46\\
14&29054155657235488&55\\
15&31426485969804308768&65\\
16&64001015704527557894928&76\\
17&245935864153532932683719776&88\\
18&1787577725145611700547878190848&100\\
19&24637809253125004524383007491432768&114\\
20&645490122795799841856164638490742749440&129\\
21&32220272899808983433502244253755283616097664&145\\
22&3070846483094144300637568517187105410586657814272&161\\
\end{tabular}
}
\end{flushleft}
where $G_{p}\approx 2^{\textrm{bits}}$ for $p=6,7,\dots ,22$, another two numbers are
{\footnotesize
$${
\begin{split}
G_{23}&=559946939699792080597976380819462179812276\\
&\quad ~348458981632\approx 2^{179},
\end{split}}$$
}
and
{\footnotesize
$${
\begin{split}
G_{24}&=1957049063020784479221748624167262560041220\\
&\quad ~75267063365754368\approx 2^{197}.
\end{split}}$$
}

\begin{flushleft}
{\footnotesize
\textbf{Table-2.} The numbers of trees of order $p$ \cite{Harary-Palmer-1973}.
\begin{tabular}{cll}
$p$&$t_p$&$T_p$\\
6&6&2\\
7&11&48\\
8&23&115\\
9&47&286\\
10&106&719\\
11&235&1,842\\
12&551&4,766\\
13&1,301&12,486\\
14&3,159&32,973\\
15&7,741&87,811\\
16&19,320&235,381\\
17&48,629&634,847\\
18&123,867&1,721,159\\
19&317,955&4,688,676\\
20&823,065&12,826,228\\
21&2,144,505&35,221,832\\
22&5,623,756&97,055,181\\
23&14,828,074&268,282,855\\
24&39,299,897&743,724,984\\
25&104,636,890&2,067,174,645\\
26&279,793,450&5,759,636,510\\
\end{tabular}
}

where $t_p$ is the number of trees of order $p$, and $T_p$ is the number of rooted trees of order $p$.
\end{flushleft}

\begin{flushleft}
{\footnotesize
\textbf{Table-3.} The numbers of digraphs and connected digraphs of order $p$ \cite{Harary-Palmer-1973}.
\begin{tabular}{cll}
$p$ & Digraphs & Connected digraphs\\
1&1&1\\
2&3&2\\
3&16&13\\
4&218&199\\
5&9,608&9,364\\
6&1,540,944&1,530,843\\
7&882,033,440&880,471,142\\
8&1,793,359,192,848&1,792,473,955,306\\
\end{tabular}
}
\end{flushleft}

\end{CJK}

\end{document}